\documentclass[symmetry,review,accept,moreauthors,pdftex]{Definitions/mdpi}
\usepackage{amsmath,amssymb}

\def\nue{\ensuremath{\nu_{e}\ }}
\def\nubare{\ensuremath{\overline{\nu}_{e}}\ }

\def\numu{\ensuremath{\nu_{\mu}\ }}
\def\nubarmu{\ensuremath{\overline{\nu}_{\mu}}}

\newcommand{\thetaot}{\ensuremath{\theta_{13}}\,}

\newcommand{\numunue}{\ensuremath{\nu_\mu \rightarrow \nu_e\,}}

\newcommand{\nubarmunubare}{\ensuremath{\overline{\nu}_\mu \rightarrow \overline{\nu}_e\,}}
\newcommand{\nova}{NO$\nu$A$\,$}

\usepackage[normalem]{ulem}
\newcommand{\stkout}[1]{\ifmmode\text{\sout{\ensuremath{#1}}}\else\sout{#1}\fi}

\firstpage{1} 
\pubvolume{1}
\issuenum{1}
\articlenumber{0}
\pubyear{2021}
\copyrightyear{2020}
\datereceived{31 July 2021} 
\dateaccepted{17 August 2021} 
\datepublished{} 
\hreflink{https://doi.org/} 
\pdfoutput=1

\Title{A New Generation of Neutrino Cross Section Experiments: Challenges and Opportunities}

\Author{A. Branca$^{1}$\orcidA{}, G. Brunetti$^{1}$\orcidB{},  A. Longhin$^{2,3}$\orcidC{}, M. Martini$^{4}$\orcidD{}, F. Pupilli$^{3}$\orcidE{}, F. Terranova $^{1}$\orcidF{} }

\AuthorNames{Antonio Branca, Giulia Brunetti, Andrea Longhin, Marco Martini, Fabio Pupilli, Francesco Terranova}

\AuthorCitation{Branca, A.; Brunetti, G.; Longhin, A., Martini, M., Pupilli, F., Terranova, F.}

\address{%
$^{1}$ \quad Dep. of Physics, Univ. of Milano Bicocca and INFN, Milano, Italy \\
$^{2}$ \quad Dep. of Physics, Univ. of Padova, Padova, Italy \\
$^{3}$ \quad INFN sezione di Padova, Padova, Italy \\
$^{4}$ \quad IPSA-DRII  Ivry-sur-Seine, France and Sorbonne Universit\'e, Universit\'e Paris Diderot, CNRS/IN2P3, Laboratoire
de Physique Nucl\'eaire et de Hautes Energies (LPNHE), Paris, France} 

\corres{Correspondence: antonio.branca@mib.infn.it}

\abstract{Our knowledge of neutrino cross sections at the GeV scale, instrumental to test CP symmetry violation in the leptonic sector, has grown substantially in the last two decades. Still, their precision and understanding are far from the standard needed in contemporary neutrino physics. Nowadays, the knowledge of the neutrino cross-section at $O(10\%)$ causes the main systematic uncertainty in oscillation experiments and jeopardizes their physics reach. In this paper, we envision the opportunities for a new generation of cross section experiments to be run in parallel with DUNE and HyperKamiokande. We identify the most prominent physics goals by looking at the theory and experimental limitations of the previous generation of experiments. We highlight the priorities in the theoretical understanding of GeV cross-sections and the experimental challenges of this new generation of facilities.}


\begin{document}

\section{Introduction}
\label{sec:introduction}
\noindent

The physics of neutrino oscillations at accelerators~\cite{Mezzetto:2020jka} has entered a new phase after the approval of DUNE~\cite{Abi:2020wmh} and HyperKamiokande (HK)~\cite{Abe:2018uyc}. These facilities have an unprecedented level of complexity, cost, and timescale, which is comparable to the most ambitious collider experiments performed in the last decades as the B-factories or the LHC experiments. Similar to collider experiments, the progress in the experimental sensitivity must be matched by the progress in theory predictions to fully reap the physics potential of the new facilities. This ideal scenario is yet to come in neutrino physics~\cite{Katori:2016yel}.

The measurement of the  size of the $\thetaot$ mixing angle~\cite{An:2012eh,Ahn:2012nd,Abe:2011fz,Abe:2013hdq} not only opened up this era but also pinned down the weakest point of the strategy to measure neutrino mixing and test the minimally extended Standard Model ($\nu$SM), i.e. the SM minimally modified to accommodate three generations of massive neutrinos.
The optimal measurement strategy is based on high-intensity neutrino beams, whose energy is between a few hundreds of MeV up to some GeV. This is one of the most critical regions for SM calculations of neutrino interactions with matter and, in particular, neutrino cross sections on moderate-$Z$ nuclei. Even more, the targets of DUNE and HK are, respectively, Argon ($Z=18$) and water ($Z=1,8$) and lay in the critical region because the mean energy of the DUNE and HK beam is 3 and 0.8~GeV, respectively. As a consequence, the standard interactions of neutrinos with matter in the most ambitious oscillation experiments under construction nowadays are known with a precision (10-30\%) that is generally inappropriate for the aim of these facilities: a 1-4\% systematic uncertainty on the measurement of the \numunue and \nubarmunubare oscillation probability~\cite{Abi:2020qib}.

DUNE and HK mitigate this risk with a very careful design of the near \cite{AbedAbud:2021hpb,Abe:2019whr}  and far \cite{Abi:2020loh,Abe:2018uyc} detectors, comparing the observed neutrino interactions rates in the two locations. Still, the time-honored near-far cancellation technique is reaching its intrinsic limitations and the systematic uncertainty currently dominates the physics reach. 

This paper addresses this issue in the most direct manner and is mirrored by the outcome of the European Strategy for Particle Physics~\cite{Strategy:2019vxc}, which has pointed out such weakness in the global effort to measure the neutrino properties. Instead of refining the near-far cancellation technique, we discuss the opportunities and challenges of a new generation of short-baseline cross section experiments,  optimized to reach percent level precision in the inclusive, differential, and doubly-differential cross-sections. These experiments must address the region of interest for DUNE and HK but should not be considered only ancillary facilities of long-baseline experiments. Neutrino interactions at the GeV scale \cite{sajjadatar2020} provide a wealth of information on the nuclear structure at moderate $Z$ and can be extended to cover the simplest nuclear system (e.g. hydrogen or deuteron) and explore the high-Z region of relevance for high-density detectors. They are the workhorse to gain a deep understanding of weak interactions in nuclear physics and provide important data for the study of nuclear structure, which can be tested against nuclear effective theories and, in the long term, lattice QCD. We call this type of facilities a \textit{new} generation of cross section experiments because major advances in this field has been achieved by the previous generation based on dedicated experiments (SciBooNE~\cite{AlcarazAunion:2009ku}, MINER$\nu$A~\cite{Rodrigues:2015hik,Aliaga:2015aqe}, WAGASCI~\cite{Giorgio:2019iuq}, ArgoNeuT~\cite{Acciarri:2020lhp}, etc.), the near detectors of long-baseline (K2K~\cite{Gran:2006jn}, MINOS~\cite{Evans:2013pka}, T2K-ND280~\cite{Abe:2014iza}, NO$\nu$A \cite{Acero:2020eit}) and short-baseline experiments (MiniBooNE~\cite{Aguilar-Arevalo:2013dva}, MicroBooNE~\cite{Abratenko:2020acr}). We expect the next generation of dedicated experiments -- running on a timescale comparable with the data-taking of DUNE and HK -- to achieve major improvements on $\nue$, $\numu$ and $\nubarmu$ cross-sections but also contribute to the short-baseline program for the search of physics beyond the $\nu$SM.

Section~\ref{sec:current_generation} 
reviews the state-of-the-art: the main achievements on cross-sections in the critical few-GeV region, the precision reached so far and prominent open issues. In Sec.~\ref{sec:future_ND}, we anticipate the improvements that can be gained by the Near Detectors of DUNE, HK and, possibly, ESS$\nu$SB~\cite{essnusb_19,Baussan:2021paj} exploiting the high-intensity, moderate precision, beams used in future long-baseline experiments: LBNF for DUNE, JPARC for HK, and the European Spallation Source for ESS$\nu$SB. High-precision beams of moderate-intensity are discussed in Sec.~\ref{sec:precision_beams} with special emphasis on  monitored neutrino beams (NP06/ENUBET~\cite{ENUBET_proposal}) and muon beams (nuSTORM~\cite{Adey:2013pio}). A review of optimized detectors for high-precision cross section measurements and their physics reach is presented in Sec.~\ref{sec:precision_detectors}. Finally, we summarize the perspectives for this new generation of experiments to be run in the DUNE/HK era in Sec.~\ref{sec:conclusions}. 

\section{The current generation of cross section experiments and their impact on neutrino oscillations}
\label{sec:current_generation}

In contemporary particle physics, neutrino cross sections play two major roles. They provide ancillary information to extract the oscillation probabilities and, at the same time, probe the nuclear media employing pure weak processes: $\nu_{e L}$ , $\nu_{\mu L}$ , $\nu_{\tau L}$ and -- if neutrinos are Dirac particles -- their right-handed (RH) antiparticles. Here, $L$ indicates the left-handed (LH) chirality component of the neutrino fields in the $\nu$SM. For the class of experiments considered here, the cross sections are the same even if neutrinos are Majorana particles, where the $\bar{\nu}_R$ states can be replaced by the RH Majorana state $\nu_R$ for each flavor ($\nu_{e R}$, $\nu_{\mu R}$, $\nu_{\tau R}$). For the sake of definiteness, we will assume Dirac neutrinos in the rest of the paper.

The ancillary role of neutrino cross section is particularly evident in oscillation experiments aimed at measuring the oscillation probability of $P_{\nu_\alpha \rightarrow \nu_\beta} (E, L)$ as a function of the neutrino energy $E$ and source-to-detector distance, the \textit{baseline} $L$. For the case of DUNE ($L\simeq 1300$~km) and HK ($L\simeq 295$~km), \numunue and \nubarmunubare are of paramount importance to establish CP violation in the neutrino sector and the number of \nue charged-current (CC) interactions observed far from the source is

\begin{equation}
    N^{e}_{FD} = \tilde{M}' \int dE \left[ \phi_{\nu_\mu}^{FD}(E) \ \epsilon_e(E) \ \sigma_{\nu_e}(E) P(\nu_\mu \rightarrow \nu_e) \ + \phi_{\nu_\mu}^{FD}(E) \epsilon_{\mu \rightarrow e} \sigma_{\nu_\mu}(E) \right]
    \label{eq:far_appearance}
\end{equation}

where $\tilde{M}'$ is the number of scattering centers (proportional to the mass and material of the detector), $\phi_{\nu_\mu}^{FD}(E)$ is the total flux integrated during the data taking at $L$ by the \textit{Far Detector} (FD), and $\epsilon_{e}(E) \sigma_{\nu_e}(E)$ is the visible cross section, i.e., the cross-section corrected by the efficiency of the detector. 
$\epsilon_{\mu \rightarrow e}$ is the misidentification probability of tagging a \numu as a $\nu_e$, e.g., due to neutral-currents (NC) events or non-reconstructed muons. For low precision experiments ($>5$\% systematic uncertainty), the uncertainties on the flux and the cross-section times detector efficiency can be removed by an identical detector located at short $L$, where oscillation effects are negligible, the \textit{Near Detector} (ND). For the ND, the \numu CC events provide an estimate of the initial flux 

\begin{equation}  
    N_{ND}= \tilde{M}_N \int dE \ \phi^{ND}_{\nu_\mu}(E) \epsilon_{\mu}(E) \sigma_{\nu_\mu}(E)
    \label{eq:near}
\end{equation}

and the \nue CC events measure the  $\nue$ contamination at source:

\begin{equation}
    N^{e}_{ND} = \tilde{M}_N \int dE \left[ \phi_{\nu_e}^{ND}(E) \ \epsilon_e(E) \ \sigma_{\nu_e}(E) \ + \phi_{\nu_\mu}^{ND}(E) \epsilon_{\mu \rightarrow e} (E) \sigma_{\nu_\mu}(E) \right].
    \label{eq:near_appearance}
\end{equation}

In this way, the main contribution to the systematic uncertainty should come only from the 
\begin{equation}
 \frac{\epsilon_e (E) \sigma_e (E)}{ \epsilon_\mu (E) \sigma_\mu (E)}    
\end{equation}
ratio~\cite{Huber:2007em} and is strongly mitigated by the lepton universality, i.e. the statement that $\sigma_e (E) = \sigma_\mu (E)$ except for calculable kinematic corrections.
Unfortunately, such an elegant solution does not work at $<5$\% level. The ND and FD are not perfectly identical. Their efficiencies are different ($\epsilon^{ND} \neq \epsilon^{FD}$) both for \nue CC and \numu CC. Even more, $\phi_{\nu_\mu}(E)$ and  $\phi_{\nu_e}(E)$ are different at the near and far location beyond the expected geometrical reduction ($\phi \sim L^{-2}$) because the ND integrates the flux in a much larger neutrino phase-space. The solid angle seen by the ND is much bigger than the FD and this phase mismatch requires corrections both on flux normalization and energy spectrum, which, in turn, increase the systematic uncertainty.  

It is worth noting that flux, efficiencies, oscillation probabilities, and cross sections are all functions of the neutrino energy $E$. The neutrino energy is not known a priori and must be reconstructed event-by-event by the final-state particles of the neutrino interaction in the detector. This reconstruction is not possible in NC events because of the outgoing neutrino and troublesome in CC events every time a final-state particle is missed or misreconstructed. Again, this procedure generates a bias affecting the measured oscillation probability.
As a consequence, assuming priors on the neutrino beam (Monte Carlo simulation of the beamline), cross-sections (neutrino-nucleus models), and detector response are mandatory even after the near-far comparison. These priors, clearly, are not free of systematic uncertainties.

Along the same line, there is little hope to predict the $\nu$SM interactions with matter starting from theory, and all advances in this field require a sophisticated interplay between model-building and high-precision cross section measurements. 

\subsection{Standard candles}

$\nu$SM provides firm predictions only for elementary fermions. For any realistic experiment at accelerators, this limitation impedes any ab-initio calculation of scatterings except for purely leptonic processes. \\

The $\numu e^- \rightarrow \mu^- \nue$ scattering is a purely leptonic \numu CC process with an outgoing $\nu_e$. The process is called \textit{inverse muon decay}. The differential cross section in $\nu$SM is:
    \begin{equation}
        \frac{d \sigma}{d \Omega} = \frac{G_F^2}{\pi^2 m_e} \frac{E_\mu^2}{E_{\nu_\mu}} \left[ E_{\nu_\mu} E_\mu  - E_{\nu_\mu} |\mathbf{k}'| \cos \theta + m_e E_\mu -m_\mu^2  \right] .
    \label{eq:numu_e_CC}
    \end{equation}
    In eq.~\ref{eq:numu_e_CC}, $G_F$ is the Fermi constant, $\mathbf{k}'$ is the three-momentum of the final-state muon measured in the laboratory frame (LAB) and $\theta$ is the corresponding scattering angle with respect to the direction of the incoming neutrino. Unfortunately, the kinematic threshold for the inverse muon decay is very large:
    \begin{equation}
        E^{thr}_{\nu_\mu} = \frac{m_\mu^2-m_e^2}{2m_e} \simeq 11 \ \mathrm{GeV}
    \end{equation}
    and this process is not particularly interesting in the DUNE/HK region. $\sigma_{\numu e^- \rightarrow \mu^- \nue} /E$ was measured at CERN and Fermilab with a precision of about 6\% up to 600~GeV~\cite{Mishra:1990yf}. \\
    
$\numu e^- \rightarrow \numu e^-$ is the threshold-less \textit{neutrino-electron scattering} corresponding to an \textit{elastic scattering} of neutrinos mediated by a $Z^0$, i.e. a neutral current. The tree-level cross section is:
\begin{gather}
    \frac{d \sigma}{d \Omega} = \frac{G_F^2}{4 \pi^2 m_e} \frac{(E'_e)^2}{E_\nu } \left[ ( g_V^e + g_A^e )^2 ( E_{\nu} E'_e  - E_\nu |\mathbf{p}'| \cos \theta + m_e E'_e -m_e^2 )  \right. \nonumber \\
    + ( g_V^e - g_A^e )^2 (E_\nu + m_e -E'_e) (E'_e - |\mathbf{p}'| \cos \theta ) \nonumber \\
    \left. - m_e \left[ (g^e_V)^2 + (g_A^e)^2 \right] (m_e -E'_e + |\mathbf{p}'| \cos \theta ) \right] . 
\end{gather}
Again, $G_F$ is the Fermi constant, $\mathbf{p}'$ is the three-momentum of the final-state electron in LAB and $\theta$ is the corresponding scattering angle with respect to the direction of the incoming neutrino.
$g_V^e = - \frac{1}{2} + 2 \cdot sin^2\theta_w$ and $g_A^e = - \frac{1}{2}$ are the $Z^0$ couplings to the electron, where $\theta_w$ is the Weinberg angle. 
This process was tested by several experiments at CERN, BNL, and Fermilab and the cross section turned out to be in good agreement with the $\nu$SM expectations. In particular, the CHARM experiment provides a total cross section of~\cite{Dorenbosch:1988is}:
    \begin{gather*}
        \sigma(\numu e^-) = 2.2 \pm 0.4 \pm 0.4 \times 10^{-42} \ E_\nu \ \frac{\mathrm{cm}^2}{\mathrm{GeV}} \\
     \sigma(\nubarmu e^-) = 1.6 \pm 0.3 \pm 0.3 \times 10^{-42} \ E_\nu \ \frac{\mathrm{cm}^2}{\mathrm{GeV}}   
    \end{gather*}
where $E_\nu$ is the neutrino energy in GeV. Note that, in the DUNE/HK region, this cross-section is $\sim$1000 times smaller than the $\nu_\mu$-nucleon cross section.
Since this process is elastic, all kinematic variables are accessible to the experimentalist if the detector can identify electrons, measure their energy $E'_e = \sqrt{ |\mathbf{p}'|^2 + m_e^2}$,  and the $\theta$ angle. The incoming neutrino direction can be considered constant ($\theta \simeq 0$) if $L$ is sufficiently large. Applying four-momentum conservation for a neutrino with four-momentum $p_{\nu} = (E_\nu,0,0,p) \simeq (p,0,0,p)$, an electron at rest in LAB ($p_e = (m_e,0,0,0)$),  and neglecting the electron mass, we get:
    \begin{equation}
        1-\cos \theta = \frac{m_e(1-T_e/E_\nu)} {E_e}
    \end{equation}
where $T_e$ is the electron kinetic energy. As a consequence, the energy and direction of the electron provide the value of $E_\nu$.
The \nue may contribute with an elastic $\nue e^- \rightarrow \nue e^-$ scattering that includes both the exchange of a $W^+$ and a $Z^0$, that is both the CC and NC contributions mentioned above. 
However, \nue are always much less than \numu (<5\%) in conventional accelerator neutrino beams and the CC contribution is negligible.

The $\nu_\mu \ e^-$ elastic scattering is the most prominent example of \textit{standard candle}, a process that is predicted with outstanding precision by the $\nu SM$ and can be used to constrain $\phi_{\nu_\mu}(E)$. A precision test of this channel to a level comparable with the theoretical predictions is out of reach at the GeV scale and is generally used at other energies to test the $\nu SM$ and perform independent measurements of the Weinberg angle. On the other hand, it is an important tool to measure the flux in very high-intensity beams~\cite{Marshall:2019vdy}.  
In the current generation of experiments, this technique has been employed by MINER$\nu$A to constrain the neutrino flux from 2 to 20~GeV accumulating 810 events. The theory uncertainty on the expected \numu flux normalization was reduced from  7.6 to 3.9\%~\cite{Valencia:2019mkf}.

\subsection{Neutrino-nucleus scattering}

Going from elementary particle scattering to neutrino-nucleon or neutrino-nucleus scattering represents a major theory challenge. The scattering of the elementary fermions (neutrino-quark) described by the $\nu$SM must be corrected by the presence of the spectator quarks, sea quarks and gluons, which impact both the formation of the final state hadrons and their reinteraction in the nuclear medium. 
In this section, we summarize the main achievements reached by current experiments: they allowed us to correct for the sizable tensions between model predictions and experimental data and shed new light on the complexity of the process. A complete ab-initio calculation would require a lattice QCD treatment that is still far-fetched, even if lattice QCD already provides interesting information on specific topics~\cite{Kronfeld:2019nfb}. On the other hand, effective models have reached a high level of sophistication boosted by a large amount of data currently available, cross-fertilization between the experimental program and model refinement, and the pressing needs of oscillation experiments at accelerators.  

The double-differential cross section describing a neutrino-nucleus scattering is~\cite{Katori:2016yel}:
\begin{equation}
\label{eq_d2s_dO_dw}
    \frac{d^2 \sigma}{d \Omega_{k'} d \omega} = \frac{G_F^2 \cos^2 \theta_C }{ 32\pi^2} \frac{|\mathbold{k}'|} {|\mathbold{k}|} L_{\mu \nu} W^{\mu \nu} (\mathbf{q},\omega)
\end{equation}
where $G_F$ is the Fermi constant. The variables describing the
\begin{equation}
    \nu_l + A \rightarrow l^- + X
    \label{eq:nu-nucleus_scattering}
\end{equation}
process are the solid angle of the final state lepton $\Omega_{k'}$, its three-momentum $\mathbf{k}'$ and the energy $\omega$ transferred by the neutrino to the nucleus. In Eq.~\ref{eq:nu-nucleus_scattering}, $l$  is a charged lepton ($e, \mu, \tau$) and $\nu_l$ the corresponding neutrino, $A$ is the target nucleus and $X$ is a generic set of outgoing particles. $\omega = E_\nu - E'_l$ is the difference between the initial-state neutrino and final-state lepton energies. We can then write the initial neutrino four-momentum as $k=(E_\nu,\mathbf{k})$, the final-state lepton four-momentum as  $k'=(E'_l,\mathbf{k'})$, and define the four-momentum transfer as $q=k-k'\equiv (\omega,\mathbf{q}$).

The $\nu$SM provides an exact expression for the leptonic tensor $L_{\mu \nu}$ because leptons are elementary fermions. For an incoming neutrino:
\begin{equation}
    L_{\mu \nu} = 8(k_\mu k'_\nu + k_\nu k'_\mu -k_\mu k^\mu -i\epsilon_{\mu \nu \alpha \beta} k^\alpha k'^\beta)
\end{equation}
while for an incoming antineutrino is
\begin{equation}
    L_{\mu \nu} = 8(k_\mu k'_\nu + k_\nu k'_\mu -k_\mu k^\mu +i\epsilon_{\mu \nu \alpha \beta} k^\alpha k'^\beta)
\end{equation}
i.e. the antineutrino leptonic tensor has a minus sign in front of the Levi-Civita tensor. On the one hand that is great news because the main cause of the neutrino-antineutrino asymmetries is located in the analytical part $L_{\mu \nu}$ of the cross section. 
This sign difference implies that the various components of the hadronic tensor weigh differently in the neutrino and antineutrino cross sections.
The name of the game is the evaluation of the hadronic tensor, which mixes up the hard weak-scattering with non-perturbative QCD and nuclear dynamics effects. The simplest process (one particle-one hole, ``1p-1h'') is the excitation that brings one  nucleon of the nucleus to a higher nuclear level leaving a hole in the nucleus ground state configuration. If the energy and momentum transfer are smaller than the production threshold of the lightest hadron (the pion), this process corresponds to a \textit{quasi-elastic} (QE) scattering where $X$ is the final state nucleus. Otherwise, $X$ is a generic set of hadrons, which generally include the final state nucleus and a few light hadrons or a real hadron cascade, the \textit{deep-inelastic} scattering (DIS). The hadronic tensor can hence be expressed  as: 
\begin{equation}
\mathbf{W}^{\mu \nu} (\mathbf{q}, \omega) = \mathbf{W}^{\mu \nu}_{1p1h} (\mathbf{q}, \omega) + \mathbf{W}^{\mu \nu} (\mathbf{q}, \omega)_{2p2h} + \mathbf{W}^{\mu \nu}_{1p1h1\pi} (\mathbf{q}, \omega)+\ldots
\label{eq:wnumu}
\end{equation}
accounting for quasi-elastic (``1p-1h'') multiple nucleon excitations (two-particles two-holes, ``2p-2h''), pion production and higher inelastic channels. According to Eq.~\ref{eq_d2s_dO_dw} the above decomposition holds for the cross section. 

The current generation of experiments has shown that the challenge of describing $\mathbf{W}^{\mu \nu}$ and the cross sections  has been overly underestimated. 
A large part of the difficulties is related to the fact that the neutrino beams are not monochromatic but wide-band. Hence, the full reconstruction of kinematics of the neutrino-nucleus reaction is impossible without assumptions. The cross sections in terms of traditional kinematic variables, like the transferred energy $\omega$ or the square of the 4-momentum transfer $Q^2=-q^2$ in the case of electron scattering, are replaced for neutrinos by the neutrino flux-integrated differential cross sections on direct observables, such as:  
 \begin{equation}
 \label{cross}
\frac{d^2 \sigma}{d\Omega_{k'}~dE'_{l}}=
\frac{1}
{ \int \Phi(E_{\nu})~d E_{\nu}}
 \int ~d E_{\nu}
\left[\frac{d^2 \sigma}{d\Omega_{k'}~d \omega  }\right]_{\omega=E_{\nu}-E'_{l}} \Phi(E_{\nu}). \end{equation}
In this type of cross sections, for fixed values of the measured variables $E'_l$ and $\Omega_{k'}$,
one explores the whole ($\omega$,$|\mathbf{q}|$) plane where the different terms of the hadronic tensor live,  hence 
all values of the energy transfer $\omega$ contribute to the cross sections.
In other words, since $E_\nu=E'_l+\omega$, \textit{for a given set of values of $E'_l$ and $\Omega_{k'}$
one explores the full energy spectrum of neutrinos} above the charged lepton energy. A crucial aspect related to this point is that all the reaction channels (quasi-elastic, 2p-2h, pion production,...)
are entangled and isolating a primary vertex process from the measurement of
neutrino flux-integrated differential cross section is much more difficult than in the cases of monochromatic (such as electron) beams, where the quasielastic and the $\Delta$-resonance bumps as well as the so-called ``dip''-region, populated by 2p-2h contributions, can be easily distinguished. 

In the following, we describe three prominent cases, where the amount and quality of new cross section data allowed us to grasp the complexity of nuclear media and inspired new interpretations, although a complete understanding is yet left to the next generation of experiments.

\subsection{Quasi-elastic region and the axial mass}
\label{sec:qe_axial_mass}

Elastic neutral currents ($\nu_l + A \rightarrow \nu_l + A$) and quasi-elastic charged currents ($\nu_l + A \rightarrow l^- + A'$), where the final state is a lepton of flavor $l=e,\mu , \tau$, are virtually the simplest neutrino-nucleus scattering. An even simpler interaction mode would be the elastic and  quasi-elastic neutrino-nucleon scattering. Even today, however, our experimental knowledge of these basic nucleon processes is limited to a set of experiments carried out with deuterium from 1981 to 1983~\cite{Miller:1982qi,Baker:1981su,Kitagaki:1983px} and early results on liquid hydrogen, plus the ancillary data from electron-nucleon scattering. The quasi-elastic cross section on nuclei plays a leading role in many neutrino oscillation experiments, especially in T2K/HK that employ water both as a target and a Cherenkov radiator and use quasi-elastic kinematics-based method to reconstruct the neutrino energy. QE are the main process 
of interest in ESS$\nu$SB~\cite{essnusb_14,essnusb_19,Baussan:2021paj} where the mean energy is just 0.2~GeV.

\begin{figure}
\begin{center}
  \includegraphics[width=6cm]{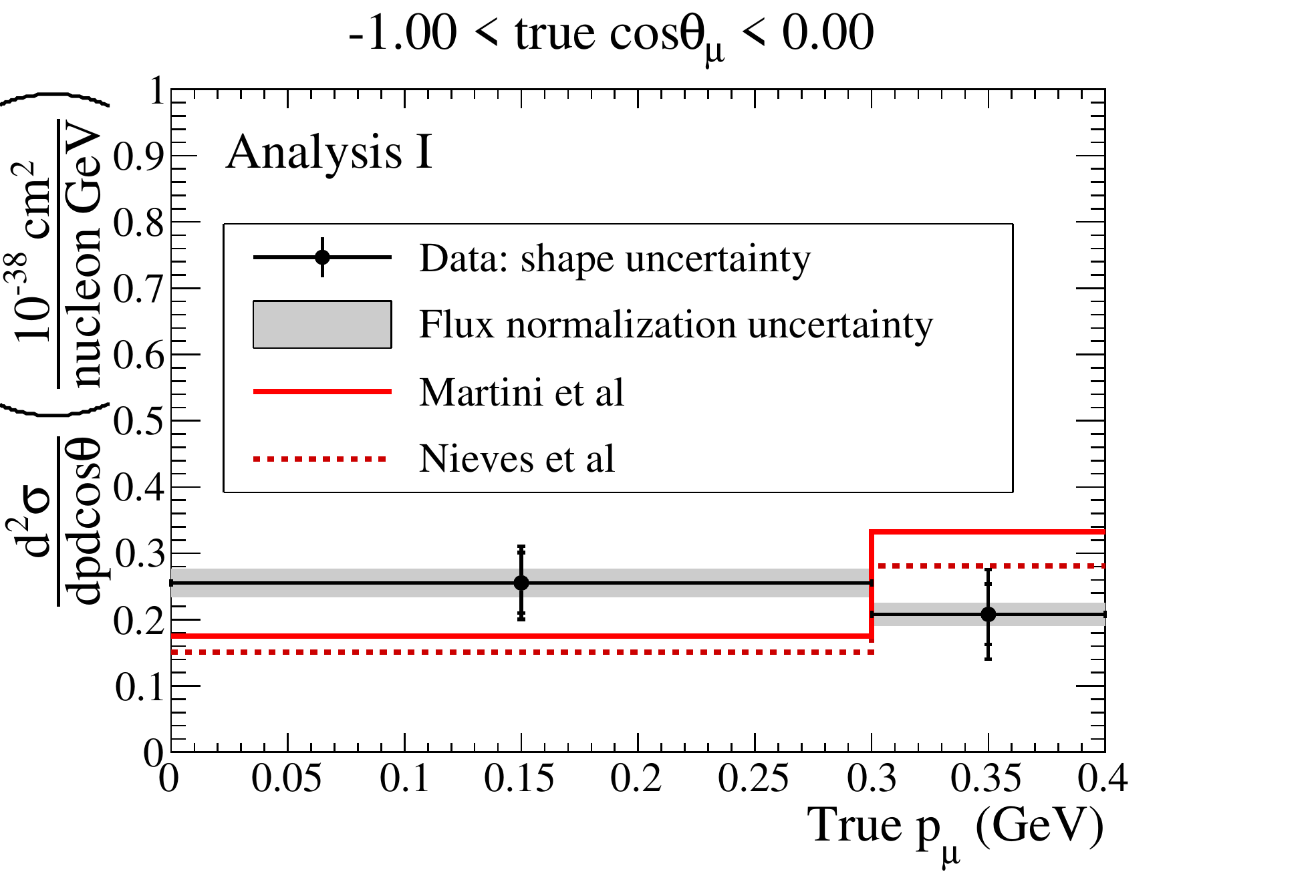}
 \includegraphics[width=6cm]{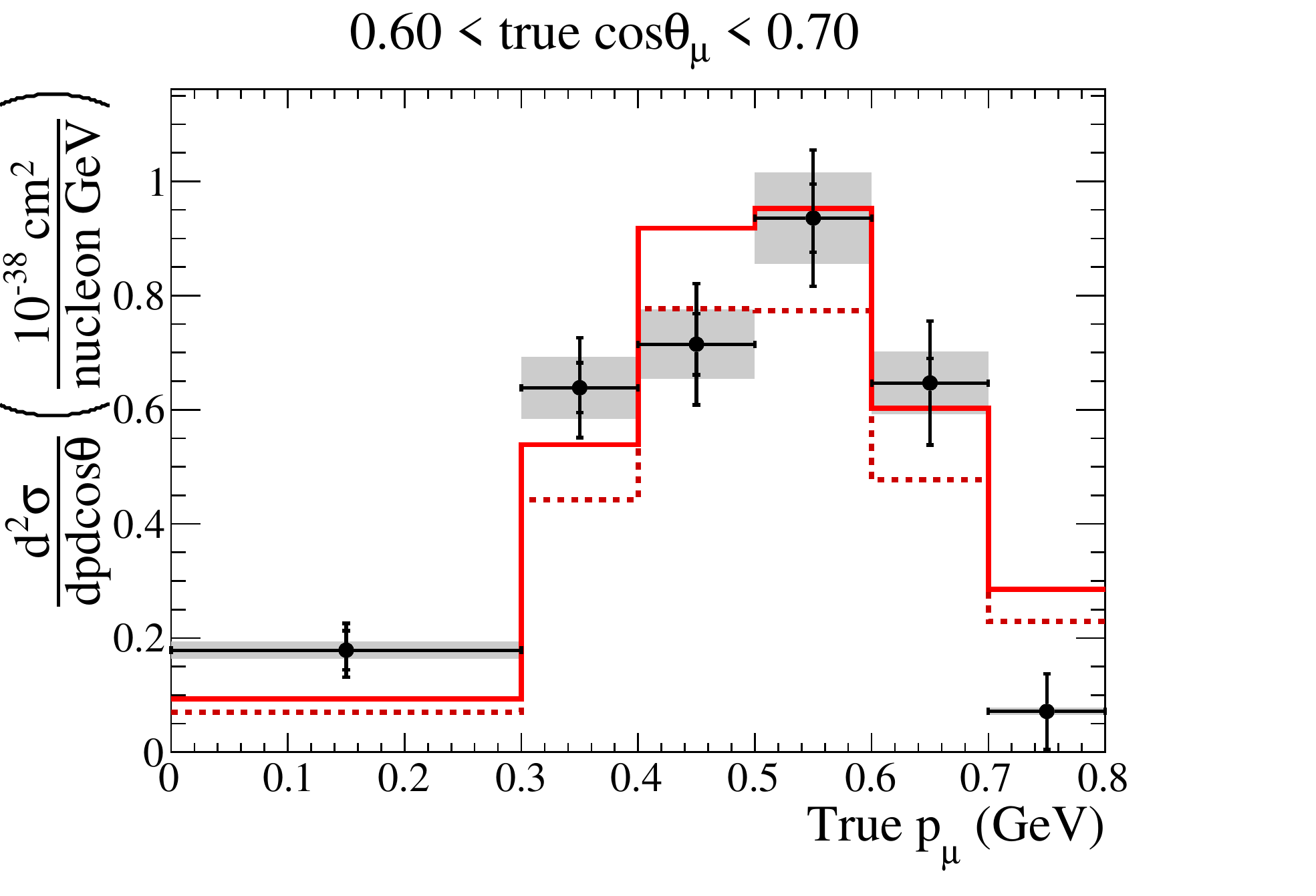}\\
 \includegraphics[width=6cm]{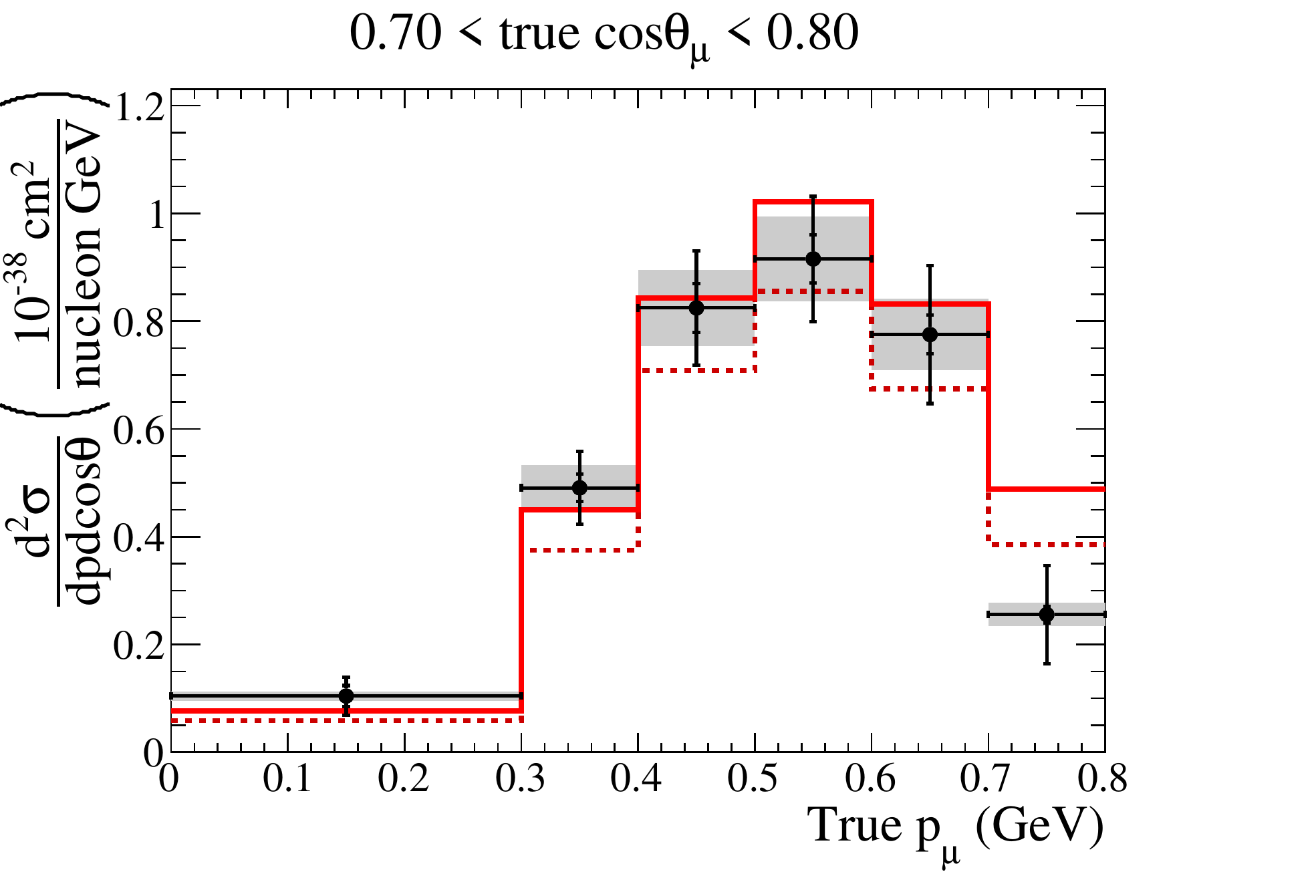}
 \includegraphics[width=6cm]{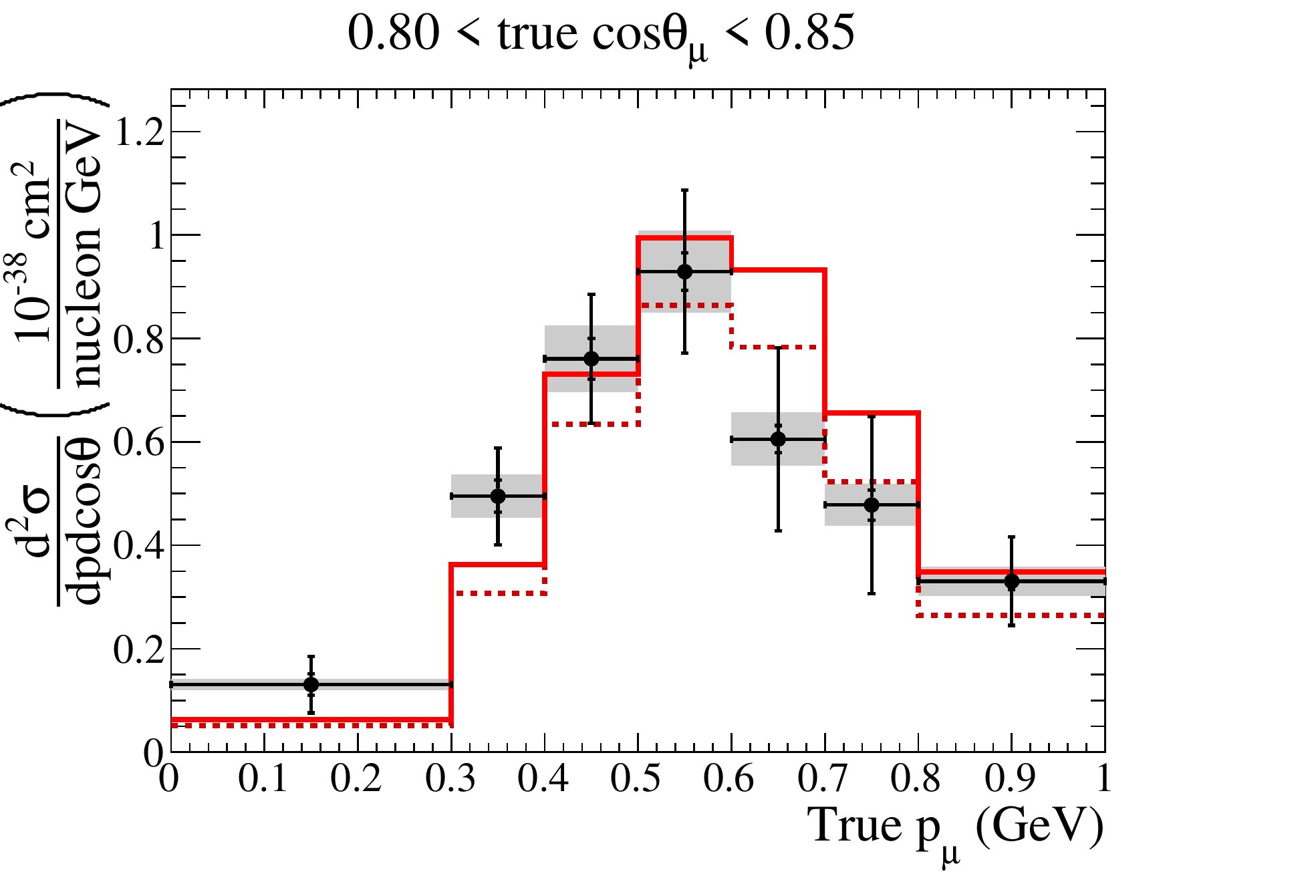}\\
 \includegraphics[width=6cm]{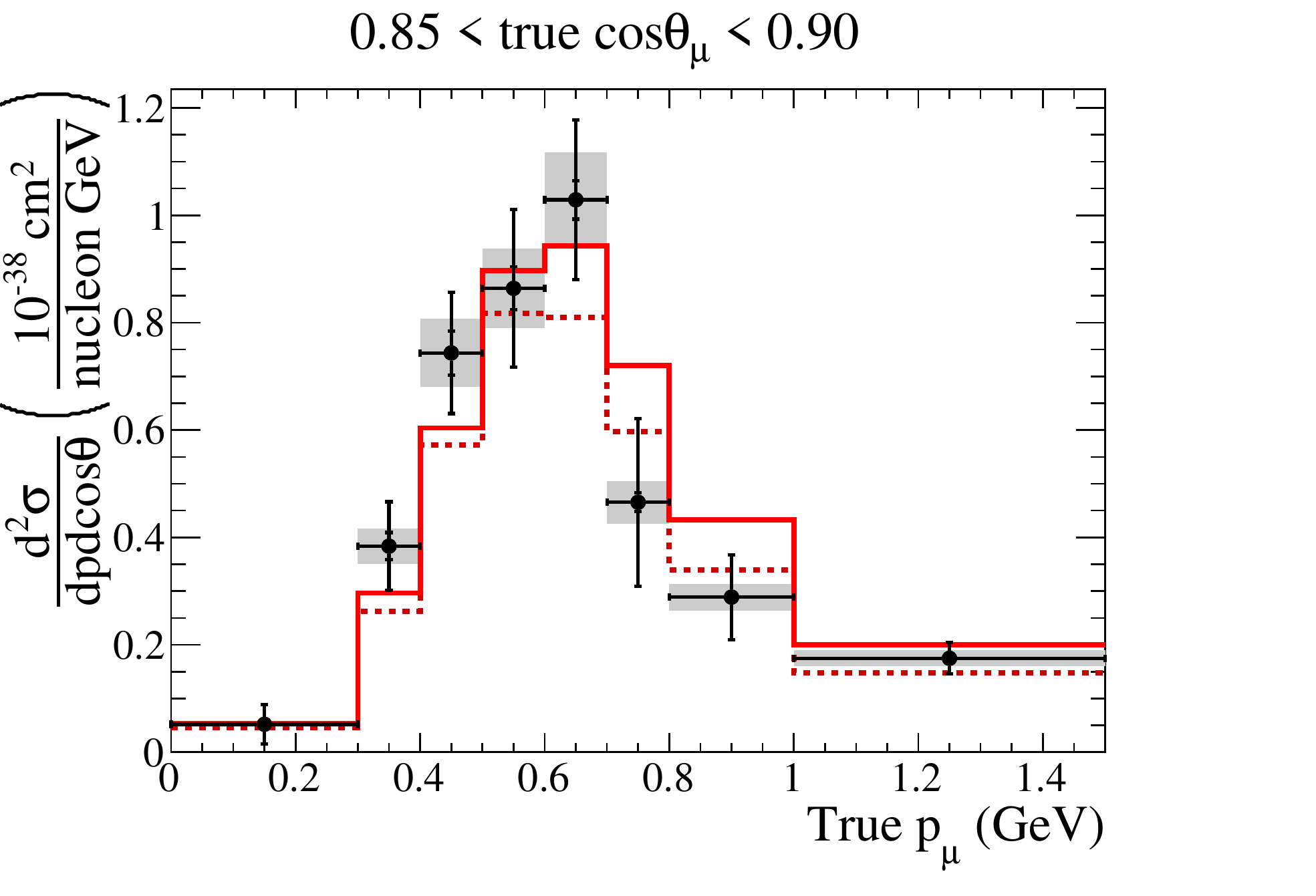}
 \includegraphics[width=6cm]{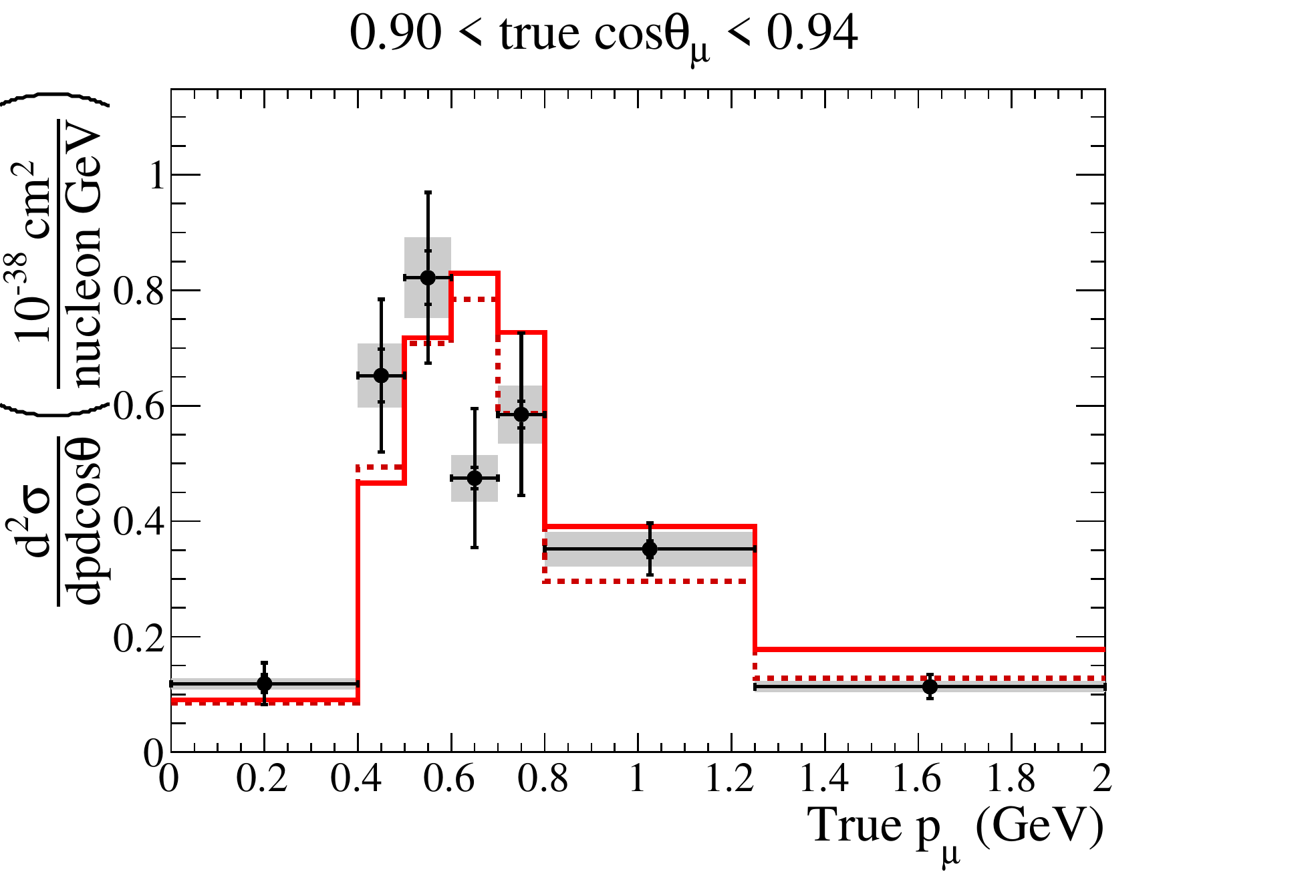}\\
 \includegraphics[width=6cm]{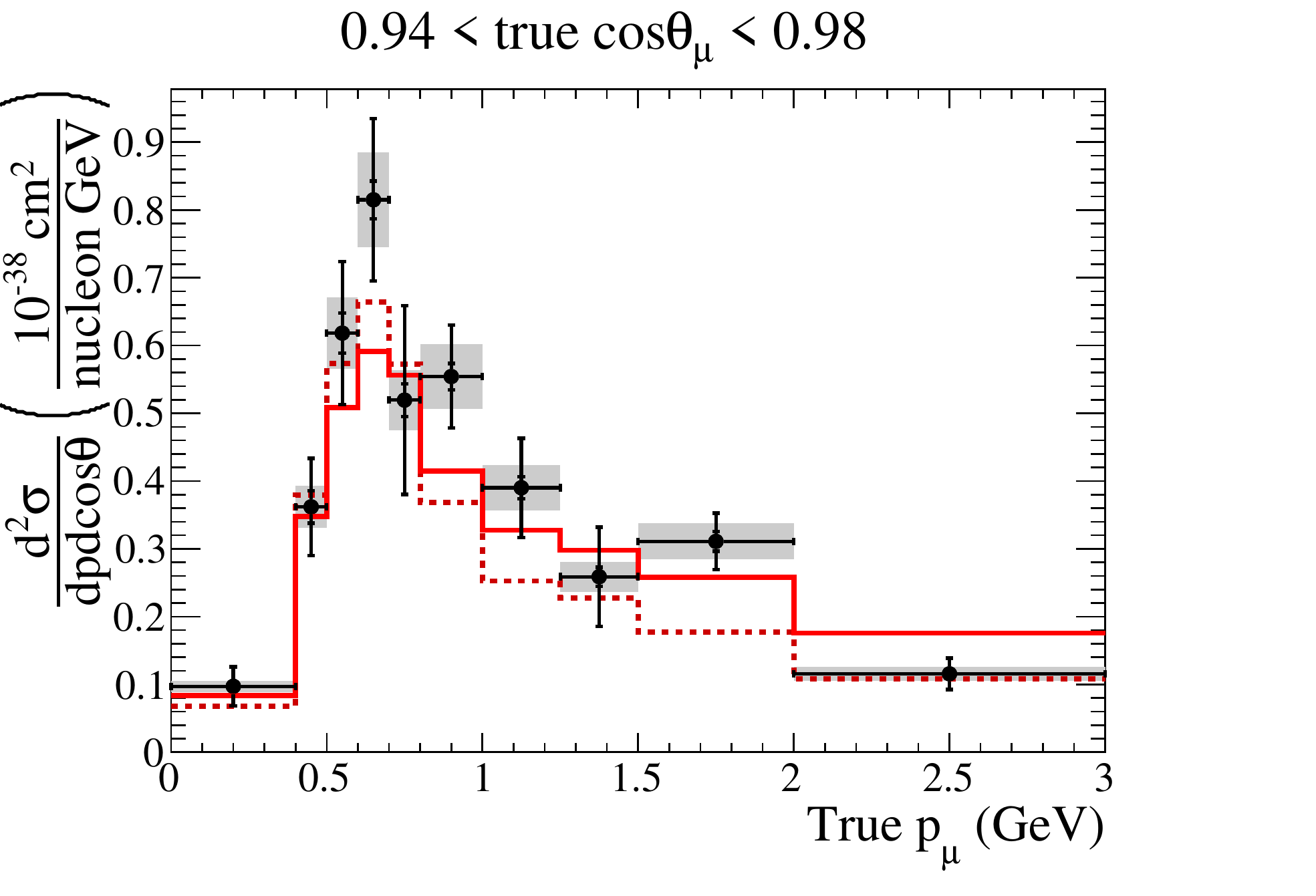}
 \includegraphics[width=6cm]{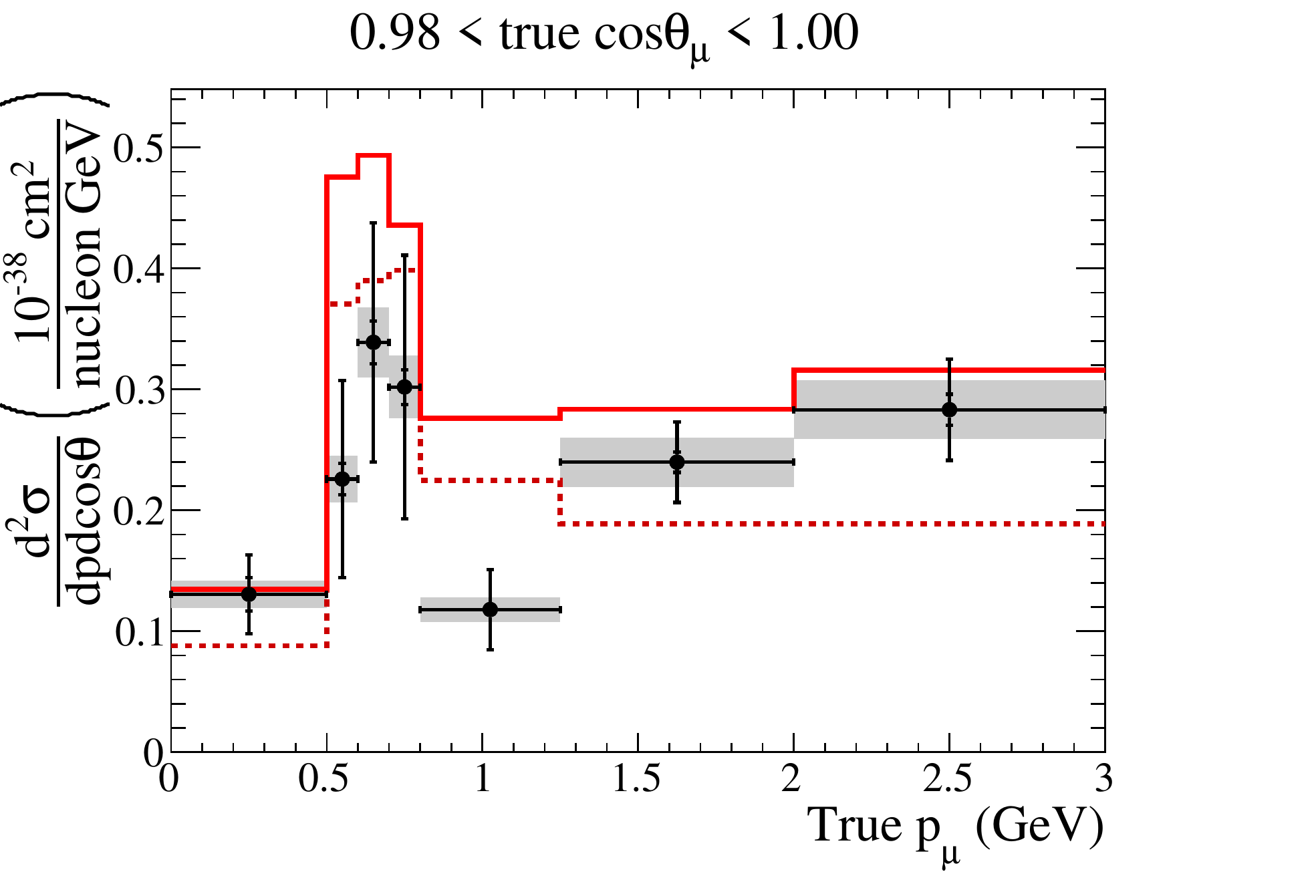}
\end{center}
\caption{The T2K $0\pi$-CC \numu double-differential cross sections measured in 2016. The flux uncertainty is shown in the gray area and the distributions are compared with the models of Martini et al.~\cite{Martini:2009uj,Martini:2010ex} (red solid line) and Nieves et al.~\cite{Nieves:2011yp,Nieves:2012yz} (red dotted line). Reproduced under CC-BY-3.0 licence from~\cite{Abe:2016tmq}. Copyright 2016 American Physical Society.}
\label{fig:double_differential}
\end{figure}

The different components of the hadronic tensor $\mathbf{W}^{\mu \nu} (\mathbf{q}, \omega)$ contain the vector and axial form factors (related to the nucleon properties) and the response functions, or structure functions, (related to the nuclear dynamics). Concerning the form factors, the conserved vector current hypothesis allows us
to apply the vector (electric and magnetic) form factors measured in electron scattering to neutrino scattering. The axial form factor instead is usually described by a dipole parameterization

 \begin{equation}
 G_A(Q^2) = G_A(0) \left( 1-\frac{Q^2}{M_A^2} \right)^{-2}  
\end{equation}

where \\ $G_A(0)\equiv G_A(Q^2\rightarrow 0)=g_A=1.2756 \pm 0.0013$ is extracted from neutron $\beta$-decay \cite{Zyla:2020zbs}.
The form factor thus runs toward the energies of interest for neutrino oscillations at accelerators (0.1-10~GeV) through an empirical constant called the axial mass $M_A$. Before precision neutrino data on nuclei were available, this value was considered quite stable around~\cite{Bernard:2001rs}
\begin{equation}
    M_A = 1.026 \pm 0.021 \ \mathrm{GeV},
\end{equation}
a value extracted 
from charged current quasielastic experiments on deuterium bubble
chambers~\cite{Mann:1973pr,Barish:1977qk,Baker:1981su,Kitagaki:1983px} and confirmed by the few-GeV data from NOMAD~\cite{Lyubushkin:2008pe} on carbon. 
However, 
modern neutrino scattering data on carbon at lower energies coming from K2K~\cite{Gran:2006jn}, SciBooNE~\cite{Nakajima:2010fp}, and MiniBooNE~\cite{AguilarArevalo:2010zc} seemed to suggest larger values of the axial mass in contradiction with the previous one. In particular, the MiniBooNE data,  
the first-ever neutrino flux-integrated double differential cross sections in terms of the measured muon variables, could be reproduced by calculating the nuclear response functions using the relativistic Fermi Gas model and increasing the axial mass to the value $M_A=1.35 (\pm 0.17)$ GeV, revealing a substantial discrepancy. The introduction of more realistic theoretical models for the nuclear response functions assuming the validity of the hypothesis that the neutrino interacts with a single nucleon in the nucleus
did not change this conclusion~\cite{AlvarezRuso:2010ia}. This contradiction brought to a reconsideration of the physical meaning of $M_A$, its interplay with the nuclear models employed to describe the nucleus, and, even more, the experimental biases. In particular, it was suggested~\cite{Martini:2009uj} that the glaring inconsistencies between models and data recorded by MiniBooNE were due to a missing component in the modeling of the cross sections: the reaction mechanism of multinucleon (2p-2h) excitations due to short-range nuclear correlations, meson exchange currents and their interference (also called one nucleon-two nucleon interference). Thus, what MiniBooNE published was not
genuine quasi-elastic data. To avoid further ambiguities on signal definition, starting from the T2K measurement of Ref.~\cite{Abe:2016tmq} (see Fig.~\ref{fig:double_differential})
data are no more classified in terms of the initial vertex of the reaction, which would require largely model-dependent background subtraction in the data analysis since the various reaction mechanisms are entangled in the neutrino cross sections.  Data are now classified in terms of final state particles, such as ``1 muon  0 pion and any number of protons'',  the so-called zero-pion CC ($0\pi$-CC) events which include also the $1\pi$-CC vertex contribution if the final pion is not detected. 
Examples of cross sections measurements with the same signal definition are the ones of Ref.~\cite{MINERvA:2018hqn} for MINERvA and of Ref.~\cite{MicroBooNE:2020akw} for MicroBooNE. 


The MiniBooNE experience also boosted the development of neutrino scattering generators that embed several (switchable) models plus the features needed by the experiments to provide mock data samples. These generators and the associated community effort~\cite{Alvarez-Ruso:2017oui} are one of the most valuable outcomes of dedicated cross section experiments.

\subsection{Pion production and the resonance region}

The region of interest for long-baseline neutrino oscillation experiments is one of the toughest from the point of view of cross section calculations. A 4-momentum transfer greater than the pion production threshold creates light mesons that may re-interact in the nuclear medium being re-absorbed or producing new particles. This region is called the \textit{resonance} (RES) region because pions are generally produced by the strong decay of nuclear resonances toward the nucleus ground state~\cite{sajjadatar2020}. The same experimental data that became available in the last two decades and challenged the evaluation of $M_A$ have unveiled the obsolescence of the semi-empirical models used by early generators in the RES region. This is particularly true for the time-honored Rein-Seghal model used as the standard reference up to $\sim$2010~\cite{Rein:1982pf}. The new approaches and models that address these items cannot be properly disentangled without high-precision flux-integrated differential cross sections. Here, the large statistics needed for single or doubly differential cross-section measurements must be deconvoluted by detector biases (unobserved pions) and uncertainties in the flux shape. These procedures can be carried on only to a limited extent. Ideally, they would require dedicated facilities where the flux is known with unprecedented precision and the detector migration matrix is extracted independently. Decoupling $\sigma_\nu \epsilon_\nu$ from $\sigma_\nu$ can be achieved e.g. using complementary detection techniques on the same beam and, possibly, the same target like, for instance, liquid argon versus gas argon detectors~\cite{AbedAbud:2021hpb}. Again, this is a goal that has inspired a new generation of experiments in the DUNE/HK era.  

\subsection{Biases in the determination of oscillation parameters}

It is unfortunate that intense accelerator-based neutrino sources are indeed sources of $\nu_\mu$ only, plus a per-cent contamination of $\nu_e$ and that the ``antineutrino'' runs are heavily polluted by $\nu_\mu$. In turn, neutrino oscillations measurements mostly rely on \numunue and \nubarmunubare transitions.
The $\nue$ total cross section is measured at $>20\%$ level~\cite{Abe:2015mxf,T2K:2014lbi,MINERvA:2015jih,T2K:2020lrr,MicroBooNE:2021ldh} and no measurements are available on \nubare.
As a consequence, the measurement of the oscillation probabilities only relies on lepton universality and takes $\sigma_{\nu_e}(E) = \sigma_{\nu_\mu}(E)$ and $\sigma_{\bar{\nu}_e}(E) = \sigma_{\bar{\nu}_\mu}(E)$ for granted except for radiative corrections arising from the electron and muon mass difference. Other phase-space effects propagate in the second-class vector current and are non-negligible in the systematic uncertainty of long-baseline experiments~\cite{Day:2012gb}. This bias is amplified by detector effects, where the reconstruction of the neutrino energy in $\nue$ CC events and the purity of the selected sample differ from the ideal case of lepton universality. These effects include differences in the neutrino spectrum, the detector response between ND and FD, the reconstruction of the neutrino energy, and the large energy spread of the neutrino beam. A quantitative study of the corresponding systematic uncertainty based on data from current facilities is detailed in \cite{FernandezMartinez:2010dm,Katori:2016yel,Alvarez-Ruso:2017oui}. Once more, designing high-precision  facilities~\cite{Charitonidis:2021qfm} to decouple $\sigma_{\nu_e}(E)$ from $\epsilon_{\nu_e} (E) \sigma_{\nu_e}(E)$ or, at least, perform a high precision measurement of $\epsilon_{\nu_e} (E) \sigma_{\nu_e}(E)$ using the same target as the long-baseline experiments (water or liquid argon) would be a remarkable asset.     

\subsection{Open issues in the theoretical understanding of cross-sections} 
\label{sec:open_issues}

\begin{figure}[t]
\begin{center}
\includegraphics[scale=0.9]{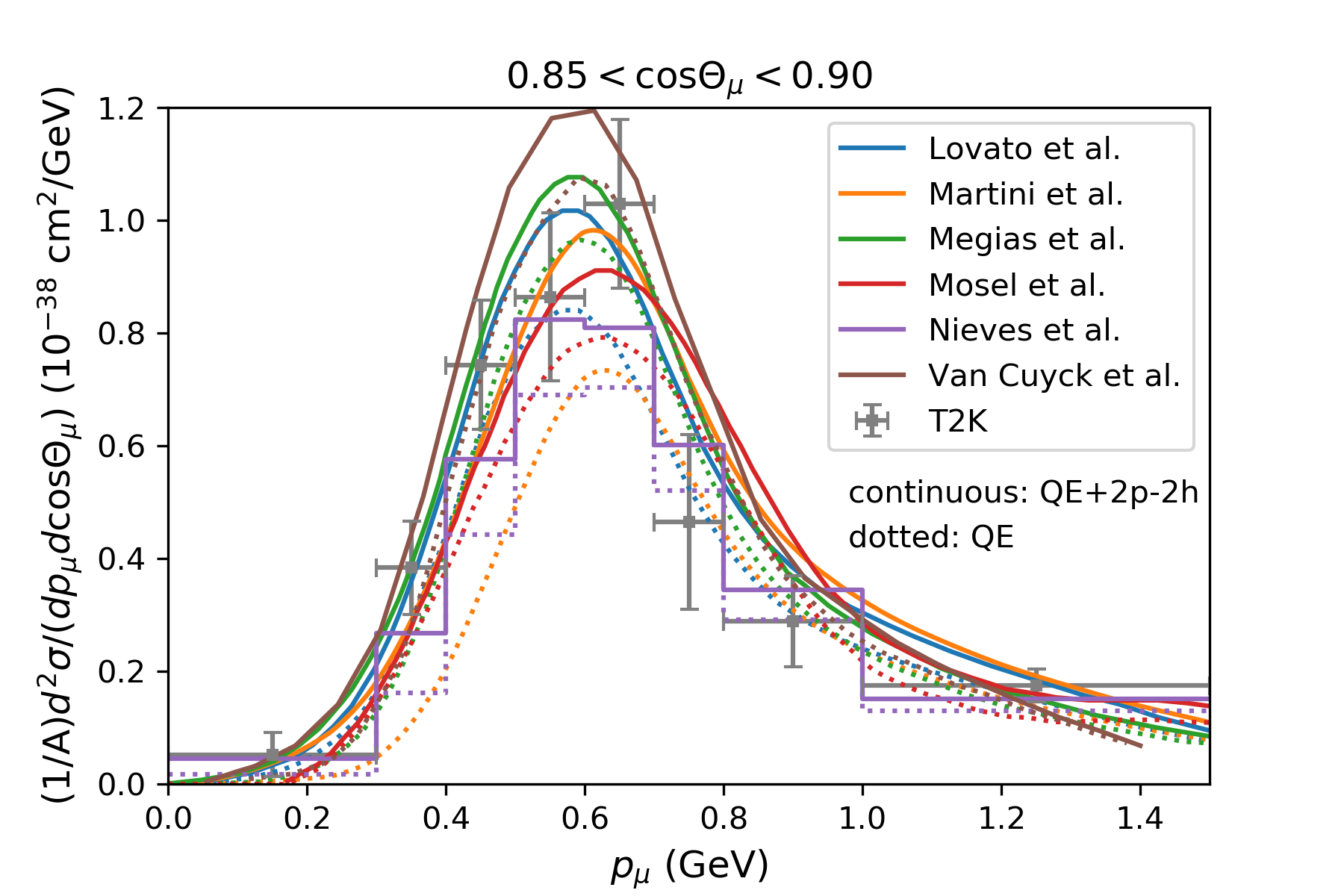}
    \caption{Comparison between different theoretical predictions of the CCQE and CCQE+2p-2h T2K flux-integrated muon neutrino double-differential cross section on carbon as a function of the muon momentum for $0.85<\cos \theta_\mu<0.90$. The experimental \textbf{CC$0\pi$} results of T2K~\cite{Abe:2016tmq} are also shown.The results of Lovato \textit{et al.} are taken from Ref.~\cite{Lovato:2020kba}. The results of of Martini~\textit{et al.}~\cite{Martini:2009uj} and Nieves~\textit{et al.}~\cite{Nieves:2011pp} appear in Ref.~\cite{Abe:2016tmq}. 
    The Megias \textit{et al.} results are taken from Ref.~\cite{Megias:2016fjk}. 
    The Mosel \textit{et al.} results are taken from Ref.~\cite{Mosel:2017anp} where the pion absorption contribution (not shown in this figure and in any case relatively small for the angle bin considered here) is also separately plotted. The Van Cuyck \textit{et al.} results are taken from Ref.~\cite{VanCuyck:2017wfn}.
    }
    \label{fig_comp_t2k_cc0pi_theory}
\end{center}
\end{figure}

As already mentioned, the MiniBooNE experience led the community to focus on neutrino flux-integrated differential cross sections in terms of the final state topology of the reactions. Furthermore, after the suggestion~\cite{Martini:2009uj} of the inclusion of 2p-2h excitations mechanism as the likely explanation of the MiniBooNE anomaly, 
the interest of the neutrino scattering and oscillation communities on the multinucleon emission channel rapidly increased. This channel was not included in the generators used for the analyses of the neutrino cross sections and oscillations experiments but turned to be crucial in the reconstruction of neutrino energy via QE-based method ~\cite{Martini:2012fa,Martini:2012uc,Nieves:2012yz,Lalakulich:2012hs,Ankowski:2016jdd}. 
The effort of the theoretical community particularly focused on this channels. Many theoretical calculations of CCQE+2p-2h and CC$0\pi$ flux integrated differential cross sections have been performed by different groups \cite{Martini:2011wp,Martini:2013sha,Martini:2014dqa,Martini:2016eec}, \cite{Nieves:2011yp,Nieves:2013fr,Bourguille:2020bvw}, \cite{Megias:2016fjk,Megias:2017cuh,Megias:2018ujz,Ivanov:2018nlm,Barbaro:2021psv},  \cite{Lalakulich:2012ac,Gallmeister:2016dnq,Mosel:2017anp}, \cite{Pandey:2014tza,VanCuyck:2016fab,VanCuyck:2017wfn}, \cite{Lovato:2020kba}. Nowadays several calculations agree on the crucial role of the multinucleon emission in order to explain the 
MiniBooNE \cite{AguilarArevalo:2010zc,AguilarArevalo:2013hm}, T2K \cite{Abe:2016tmq,T2K:2017qxv,T2K:2019ddy,Abe:2020jbf,T2K:2020jav}, MINERvA \cite{MINERvA:2018hqn,MINERvA:2019gsf} and MicroBooNE \cite{MicroBooNE:2020akw} cross sections. Nevertheless there are some differences on the results obtained by the different theoretical approaches. An illustration of the amount of the differences between the results obtained by different theoretical approaches for the CCQE and the CCQE+2p-2h
is given in Fig.~\ref{fig_comp_t2k_cc0pi_theory} where the T2K flux-integrated double-differential cross section on carbon is shown as a function of the muon momentum for $0.85<\cos \theta_\mu<0.90$ and it is compared to the experimental \textbf{CC$0\pi$} T2K results, already introduced in Fig.~\ref{fig:double_differential}.
At the present level of experimental accuracy quantifying
the agreement between the T2K data and the different models is not evident; the uncertainties are too large for any conclusive statement. However, even if a one-to-one correspondence between the different exclusive channel
contributions of different theoretical calculations is not always possible (for example short-range correlations are considered as part of genuine QE in the approaches of Refs.~\cite{Megias:2016fjk},\cite{Lovato:2020kba}, while they are separately calculated as part of the 2p-2h excitations in the case of  Refs.~\cite{Martini:2009uj},\cite{Nieves:2011pp},\cite{Mosel:2017anp},\cite{VanCuyck:2017wfn}), we can consider Fig.~\ref{fig_comp_t2k_cc0pi_theory} as an illustration of the fact that also T2K data seems to
prefer CCQE+2p-2h with respect to pure CCQE calculations.

A more complete comparison between different theoretical predictions, Monte Carlo \cite{Andreopoulos:2009rq,Hayato:2009zz,Golan:2012wx}, and the data is published in Ref.~\cite{Abe:2020jbf}. In this work, the T2K flux-integrated double differential $\textrm{CC0}\pi$ cross sections, for muon neutrino and antineutrino are given, as well as their combinations important for CP-violation investigations. 
From Fig.~\ref{fig_comp_t2k_cc0pi_theory} and Ref.~\cite{Abe:2020jbf} it turns out that although the trends of the theoretical calculations are similar, some significant differences remain, including even differences between calculations of the same category, such as the ones of Martini~\textit{et al.}~\cite{Martini:2009uj},  Nieves~\textit{et al.}~\cite{Nieves:2011pp}. These differences are illustrated for instance in Ref.\cite{Katori:2016yel}, where a comparison is shown between data and predictions of models which calculate several neutrino and antineutrino MiniBooNE, T2K, and MINERvA flux-integrated differential cross sections. Moreover, the approximations made by the different groups in the treatment of 2p-2h excitations are also discussed in this work.

A particularly important region is the one of very forward muon scattering angle. Here not only the different theoretical models and Monte Carlo show differences among them but also they globally overestimate the data. This is the case of the $\textrm{CC0}\pi$ T2K cross sections~\cite{Abe:2016tmq,Abe:2020jbf} (as it appears for example in the $0.98<\cos \theta_\mu<1.00$ panel of Fig.~\ref{fig:double_differential}), of the T2K CC inclusive cross sections of Ref.~\cite{Abe:2018uhf}, characterized by an increased angular acceptance and higher statistics with respect to the ones of Ref.~\cite{Abe:2013jth}, and of the MicroBooNE CC1p0$\pi$~\cite{Abratenko:2020acr} and CC0$\pi$~\cite{MicroBooNE:2020akw} cross sections on argon. A deeper understanding of the very forward region is important for CP-violation experiments and should be further investigated. For example, by analyzing neutrino scattering in the giant-resonances and quasi-elastic region for fixed values of neutrino energy, it has been raised~\cite{Martini:2016eec} and investigated ~\cite{Ankowski:2017yvm,Nikolakopoulos:2019qcr,Gonzalez-Jimenez:2019qhq} the surprisingly dominance of the $\nu_\mu$ cross sections over the $\nu_e$ ones for small scattering angles. 

Several reasons contribute to the difficulty of a precise modeling of the flux integrated differential cross sections in the very forward region: 
\begin{itemize}
    \item The Pauli blocking effects play a significant role \cite{Megias:2017cuh,Ivanov:2018nlm,Gonzalez-Jimenez:2019qhq}
    \item The giant resonances, neglected by many theoretical models, contribute to the cross section \cite{Pandey:2016jju}
    \item The neutrino cross sections, in general dominated by the spin-transverse response, becomes sensitive to the spin-longitudinal one which is characterized not only by low-energy giant resonances but also by high-energy collective states such as  coherent mixture of $\Delta$-hole states and pions \cite{Delorme:1985ps,Martini:2014dqa}
    \item Even in neutrino beams, such as T2K/HK and MiniBooNE/MicroBooNE, which induce nuclear excitations dominated by quasielatic, 2p-2h and $1\pi$ production, higher energy excitations up to the DIS contribute, as illustrated for example in Refs.~\cite{Gallmeister:2016dnq,Bourguille:2020bvw}.
\end{itemize}
The ultimate goal should be the development of a theoretical unified approach for the treatment of all nuclear excitations, from low-energy giant resonances to DIS. For the moment flux-integrated differential cross sections calculations are available from QE to DIS in the case of the GiBUU \cite{Buss:2011mx} implementation of the transport theory \cite{Lalakulich:2012ac,Gallmeister:2016dnq,Mosel:2017anp} and from and from QE to 1 pion production induced by $\Delta$-resonance excitations in the approaches of Martini \textit{et al.} \cite{Martini:2009uj,Martini:2011wp,Martini:2014dqa}, Nieves \textit{et al.} \cite{Nieves:2011pp,Nieves:2011yp,Bourguille:2020bvw} and SuSA/SuSAv2 \cite{Ivanov:2012fm,Ivanov:2015aya,Megias:2016fjk,Amaro:2019zos,Barbaro:2021psv}. Promising approaches exploiting the factorisation of the nuclear cross section, based on spectral functions \cite{Rocco:2015cil,Vagnoni:2017hll,Rocco:2018mwt,Rocco:2019gfb} and superscaling \cite{Amaro:2019zos,Barbaro:2021psv} have already been extended to inelastic region beyond $\Delta$-resonance and employed to investigate electron scattering and neutrino cross sections for fixed kinematics and as a function of the neutrino energy. Neutrino flux-integrated differential cross sections represent the next step. 

The developed theoretical nuclear models
should not only cover different reaction channels and kinematics but also 
be applicable to various nuclear targets.
If many calculations exist for the doubly magic $N=Z$ nuclei $^{12}$C and $^{16}$O, the effort to investigate the $^{40}$Ar started only recently \cite{Meucci:2013gja,Gallmeister:2016dnq,VanDessel:2017ery,Akbar:2017dih,Barbaro:2018kxa,Barbieri:2019ual,VanDessel:2019atx,VanDessel:2019obk,Butkevich:2021sfn,Franco-Patino:2021yhd,Barbaro:2021psv} and often it requires formal generalization of the existing approaches, being the $^{40}$Ar an open-shell nucleus with $N\neq Z$.

Precise predictions and measurements of hadronic final states represent another very important next
step. The community is moving toward to this path. 
Beyond the inclusive cross section, largely discussed above, where only the outgoing lepton is detected, there is a rapidly increasing interest on the semi-inclusive cross section, which corresponds to detecting in coincidence the outgoing lepton and an hadron (or a system of hadrons), and on the exclusive cross section, corresponding to the simultaneous detection of all the final scattering products. The last two processes are more challenging from both experimental and theoretical points of view. In the recent years the community has focused in particular on semi-inclusive reactions where a muon and one proton are detected in the final state since they can be used to identify relevant nuclear effects related to both the initial state dynamics and to final state interactions, as well as to 2p-2h excitations. Special attention has been paid to the study of kinematical variables obtained by projecting the final lepton and the ejected nucleon momenta on the plane transverse to the neutrino beam. 
The cross sections in terms of these observables, called Single-Transverse (Kinematic) Variables (STVs) or transverse kinematic imbalances (TKI) \cite{Lu:2015tcr}, are useful to enhance some nuclear effects, and therefore discriminate between different models. T2K \cite{T2K:2018rnz} and MINERvA 
\cite{MINERvA:2018hba,MINERvA:2019ope} measurements have been published in terms of these variables. Semi-inclusive CC1p0$\pi$ cross sections function of muon and proton kinematics have been published also by MicroBooNE \cite{MicroBooNE:2020akw,MicroBooNE:2020fxd}. The theoretical investigation on semi-inclusive cross sections \cite{Moreno:2014kia,VanCuyck:2016fab,VanCuyck:2017wfn,VanOrden:2019krz,Sobczyk:2020dkn,Franco-Patino:2020ewa,Gonzalez-Jimenez:2021ohu}
and on transverse kinematic imbalances \cite{Dolan:2018sbb,Bourguille:2020bvw,Franco-Patino:2021yhd} are only at the beginning, as well as the studies related to the emission of nucleon pairs induced by meson exchange currents and short range correlations~\cite{VanCuyck:2016fab,VanCuyck:2017wfn,RuizSimo:2016ikw,Pastore:2019urn}. 

Semi-inclusive processes characterized by the detection of one muon and one pion in the final state deserve also further deep investigations. In this case data-theory agreement remains very unsatisfactory.  
Nowadays there is no model which can describe MiniBooNE \cite{AguilarArevalo:2009ww,AguilarArevalo:2010bm},
MINERvA \cite{Eberly:2014mra,McGivern:2016bwh} and T2K \cite{Abe:2016aoo} data simultaneously. For example the model of Refs.\cite{Hernandez:2013jka,Alvarez-Ruso:2020bxl} allows to get a reasonable description of MiniBooNE data but over-predicts the MINERvA ones and lacks forward pions in the case of T2K. On the contrary, 
the models of Refs.\cite{Lalakulich:2012cj,Lalakulich:2013iaa,Mosel:2015tja} and Refs.\cite{Gonzalez-Jimenez:2017fea,Nikolakopoulos:2018gtf} reproduce well the shape and
strength of the MINERvA and T2K data but underpredict the MiniBooNE ones. This is the so called ``pion puzzle''.
The complications of pion data analyses lay not only on their primary production models,
but also on the pion final state interactions and on the fact that all hadronic processes related to shallow inelastic scattering (SIS) and DIS regions have to be modeled correctly, another major challenge \cite{Katori:2016yel,Alvarez-Ruso:2017oui,SajjadAthar:2020nvy,Alvarez-Ruso:2020ezu}.

\section{Learning from the next generation of near detectors}
\label{sec:future_ND}


The near detectors of long-baseline experiments like \nova and T2K have been instrumental to improve our knowledge of cross-sections and unveiled a complexity that was not suspected at the time of the discovery of neutrino oscillations. The new generation of near detectors for DUNE and HK will exploit mega-watt class  beams even if the level of control of these beams (diagnostic, flux, flavor at source, \nue and wrong-sign contamination, etc.) will be comparable to their predecessors. Since systematics is the main limiting factor to CP sensitivity both in DUNE and HK, the corresponding collaborations are devising a near detector complex much more complete than what has been done in the past. It is worth stressing, however, that the main aim of these detectors is to minimize the uncertainty on the measurement of unoscillated \numu and \nue interactions, that is the difference in the number of expected events if no oscillation occurs. If the near and far unoscillated rates were identical, the near-far difference as a function of the energy would provide a measurement of the oscillation probability free of systematic uncertainties.  
This goal is not equivalent to cross-section measurements even if cross sections play a major role. Cross section measurements are thus a byproduct of the DUNE/HK systematic mitigation strategy and reap the lessons learned by \nova and T2K.

For the sake of illustration, we detail what the DUNE near detector (ND) complex \cite{AbedAbud:2021hpb} can offer to the physics of neutrino cross section, identify its strength and point out to possible limitations.
Similar considerations hold in a narrower energy range for HyperKamiokande.

\begin{figure}[t]
\begin{center}
\includegraphics[scale=0.15]{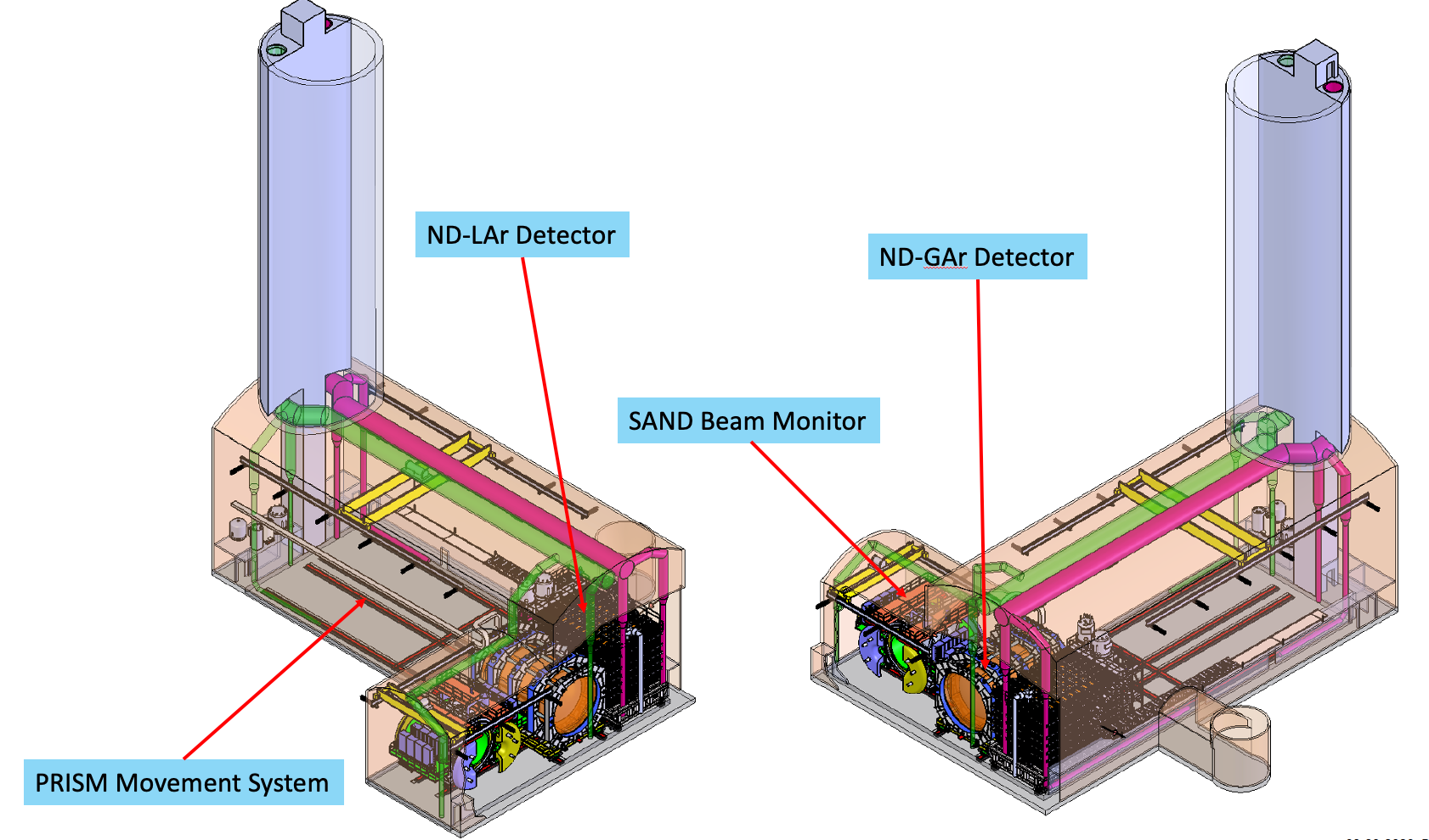}%
\includegraphics[scale=0.2]{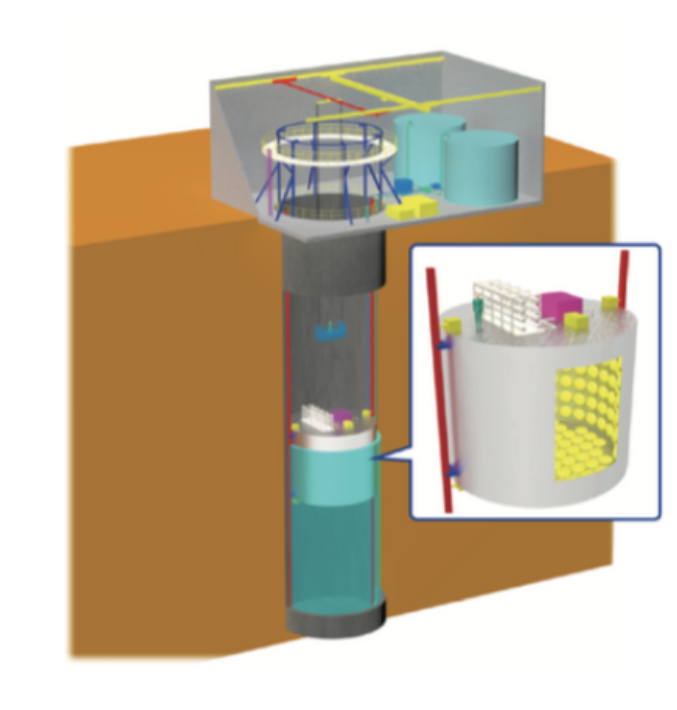}
\caption{Left: Overview of the  DUNE near detector complex with the gas argon TPC replacing the TMS. Right: the HyperKamiokande Intermediate Water Cherenkov Detector.}
\label{fig:dune_nd}
\end{center}
\end{figure}

The DUNE ND complex is shown in Fig.~\ref{fig:dune_nd} and it is made up of three subsystems:

\begin{itemize}
    \item SAND: an on-axis detector based on the KLOE magnet and calorimeter complemented by a low-density tracker 
    \item ND-LAr: a non-magnetized liquid argon TPC capable to stand the high neutrino rate of the DUNE beam 
    \item TMS: a muon catcher that closes the kinematics of \numu CC events in the liquid argon. In a second stage, this muon catcher will be replaced by a magnetized high-pressure argon TPC (ND-GAr) inside a superconducting solenoid 
\end{itemize}

The expected rates at ND-LAr is $10^8$ \numu CC and about $2 \times 10^6$ \nue CC per year. Similarly, the expected rate in SAND is $10^7$ \numu CC and $\simeq 2 \times 10^5$ \nue CC per year. 
The liquid argon TPC and the muon catcher (or the gas argon TPC) are movable detectors that run in the horizontal plane perpendicular to the beam. In this way, DUNE can monitor its beam both on-axis and off-axis. Unlike current systems like the on-axis INGRID detector and the off-axis ND280 detector at T2K, the off-axis detectors can span from 0 m (on-axis) to 30 m far from the beam axis. They can thus implement the PRISM technique \cite{Bhadra:2014oma,AbedAbud:2021hpb} to sample the neutrino flux at several angles and, hence, at several energies. The combination of these data provides a faithful representation of the unoscillated $\phi_{\nu_\mu}(E) \sigma_{\nu_\mu}(E) \epsilon_{\nu_\mu}(E)$ and $\phi_{\nu_e}(E) \sigma_{\nu_e}(E) \epsilon_{\nu_e}(E)$ at the far detector in South Dakota. This strategy is also envisaged in HK with a water Cherenkov detector located at about 1~km from the source and moving in the vertical plane perpendicular to the beam axis (1-4$^\circ$) \cite{Bhadra:2014oma} as shown in Fig.~\ref{fig:dune_nd}, right. 

\begin{figure}[t]
\begin{center}
\includegraphics[scale=0.5]{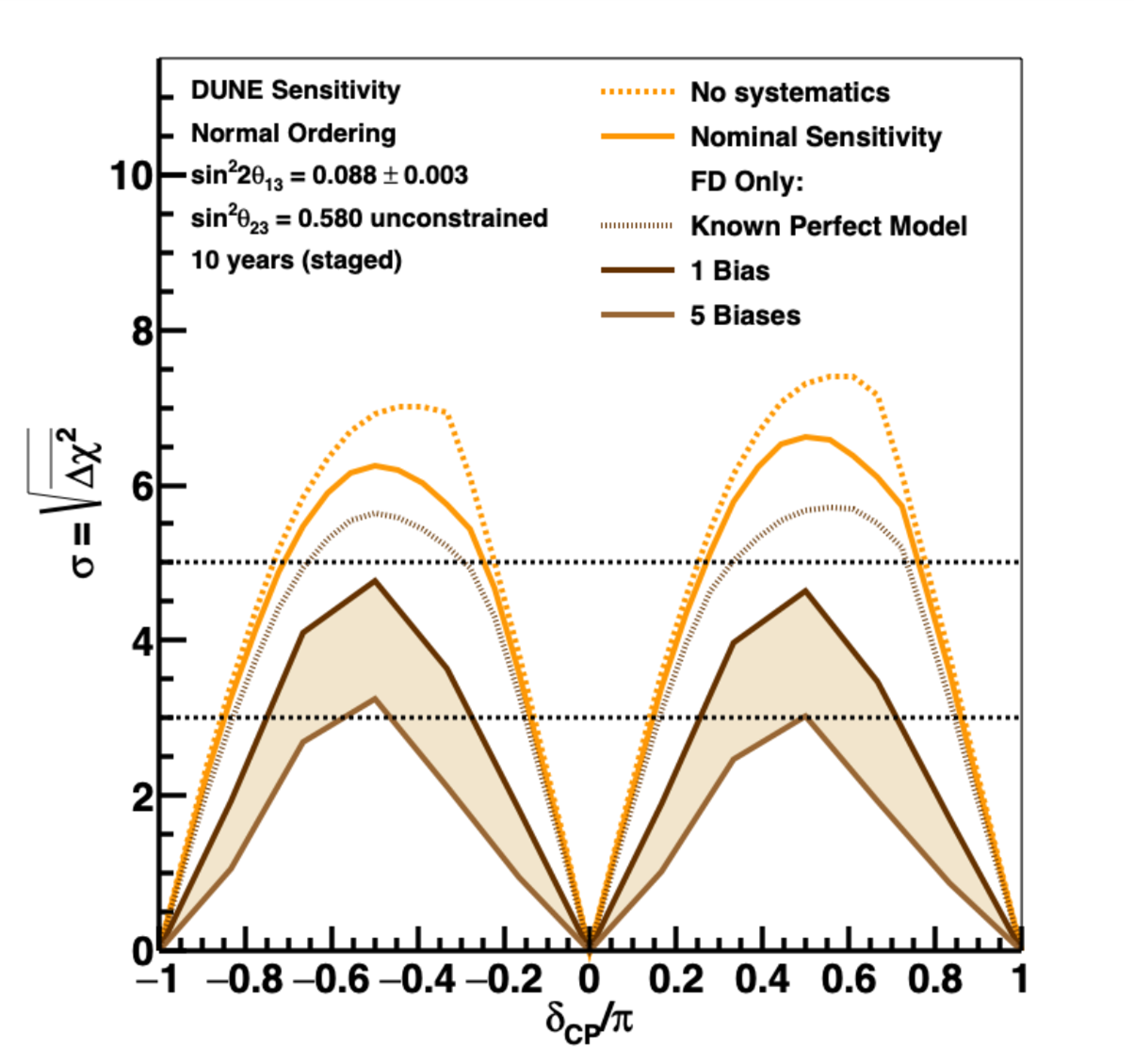}
    \caption{Impact on the DUNE CP sensitivity from uncertainties in the neutrino interaction models.}
    \label{fig:impact_dune}
  \end{center}
\end{figure}

The triple product flux$\times$cross-section$\times$efficiency in multiple energy bins is the main observable to cope with the DUNE and HK systematics due to the near-versus-far difference.
The triple product can be partially deconvolved to attain information on the cross sections. The main tool for deconvolution is an independent measurement of the neutrino flux. DUNE and HK are planning to exploit the aforementioned standard candles: the neutrino-electron scattering and the low-$\nu$ events~\cite{Belusevic:1988ab}. In the most aggressive scenario, standard candles combined with the classical techniques to reduce the \numu flux systematics (beam simulation and hadroproduction) might achieve a precision of 2-3\%, especially in the low-density detectors: SAND and the gas argon TPC in DUNE, and the upgraded ND280 in HK.
The \nue flux systematics cannot be estimated with standard candles due to the low flux of \nue with respect to $\nu_\mu$. Its estimated uncertainty ranges from 7\% to 10\%, depending on whether DUNE runs in neutrino or antineutrino mode. Unfortunately, the PRISM technique is not enough to cancel the near-versus-far systematics even if it is an important mitigation tool. The main obstruction is the neutrino energy reconstruction that is still obtained by the reconstructed final state particles in the neutrino detector. Citing Ref. \cite{AbedAbud:2021hpb}, PRISM ``will minimize uncertainties in the extracted oscillation parameters arising from the spectral differences as implemented in the imperfect interaction (cross section) model. ... DUNE will need
a vibrant program of cross section measurements as input to that work''. The near detectors themselves can perform some of these measurements comparing the double-differential cross sections measured in the liquid argon with theory expectations and tuning the corresponding Monte Carlo codes. Fig.~\ref{fig:impact_dune} shows the impact on the CP sensitivity assuming modest (1 bias) and  
conservative (5 bias) biases in the interaction models \cite{AbedAbud:2021hpb}.

Decoupling the detector efficiency $\epsilon_{\nu_\mu}(E)$ and $\epsilon_{\nu_e}(E)$ from the corresponding cross section in the triple product is a major challenge. The most direct approach is measuring  the same cross section with a high-density and low-density detector simultaneously. Ideally, the medium should be the same to avoid additional biases due to the $Z$ dependence of nuclear reinteraction effects. In DUNE, this is the rationale of using simultaneously a liquid argon and gas argon TPC, provided that the flux decoupling is performed at the per-cent level. A compromise is to use detectors with the same target nuclei interspersed by high-precision trackers. This is done in HK combining the measurement on carbon and oxygen (ND280 and WAGASCI) with the measurement in water (IWCD, NINJIA, ND280-FGD2 water layers). In DUNE, besides the gas/liquid argon TPC, the SAND and liquid argon TPC data can be combined, as well. At present, however, none of these studies have been performed at percent level precision.

Tuning the neutrino interaction models with data is not equivalent to building reliable models, which should be based on low Z (hydrogen or deuteron) targets and scaled up to argon and oxygen. No near detector is equipped with these devices but SAND could perform some of these studies looking at the transverse momentum imbalance and exclusive state reconstruction, 
already introduced in Sec. \ref{sec:open_issues}, to isolate samples of events enriched in interactions on hydrogen within a hydrocarbon
target. This technique has been proposed by several authors \cite{Lu:2015tcr,Duyang:2019prb,Munteanu:2019llq,Hamacher-Baumann:2020ogq} and recently used with the T2K near detector data in \cite{PhysRevD.103.112009}. The systematic uncertainties associated with the subtraction of the non-hydrogen components of the target (mostly carbon) are under evaluation.

\section{A new generation of neutrino beams}
\label{sec:precision_beams}

As discussed in Sec.~\ref{sec:future_ND}, the disentangling of the flux, cross-section, and efficiency triple product is not fully accomplished either by current experiments or by the forthcoming near detectors and deserves a dedicated facility. This facility will be the core of a new generation of cross section experiments capable to provide firm guidance to model builders instead of tuning approximate models with data.  

The first hindrance toward high-precision cross section measurements is the uncertainty on flux. This is a prominent motivation for a new generation of short-baseline beams, which can provide controlled fluxes at a level unattainable in the near detectors of Sec.~\ref{sec:future_ND}.   
The MINER$\nu$A experiment performed the best measurement of the flux in the 1-10 GeV range on the NuMI beam~\cite{numi:flux} and showed that classical tools like GEANT4/FLUKA simulations combined with beam and muon monitors can achieve a precision of 10-25\% depending on the energy. By including hadro-production data the uncertainty can be reduced to 5.4\%.
By using an independent measurement as the $\nu_\mu \ e^{-}$ scattering for the integrated flux the precision of cross-section measurements in MINER$\nu$A can be further improved: the experiment was able to reduce such uncertainty from 7.5\% to 3.9\% by using the $\nu_{\mu} + e^{-} \rightarrow \nu_{\mu} + e^{-}$ scattering which is characterized by a purely leptonic cross section known with great precision~\cite{Valencia:2019mkf}.
This technique can be used in DUNE and HK as well. However, it will reach its intrinsic limitation (detector systematics on $\nu_\mu \ e^-$ scattering and hadroproduction data) most likely at 3\% level and will not provide information on the $\nu_{e}$ flux since their high-intensity neutrino beams are almost completely made of $\nu_{\mu}$.

\subsection{Monitored neutrino beams}
A facility like the one envisaged by the NP06/ENUBET project addresses directly this problem. It aims to monitor a neutrino beam by measuring associated charged leptons produced in a narrow-band meson beam. Large-angle leptons from kaon decays are recorded in an instrumented decay tunnel while the neutrino travels toward the neutrino detector. Since we produce a neutrino for each lepton in the tunnel, lepton counting allows for a direct flux prediction that bypasses the uncertainties coming from hadro-production yields in the targets, beamline simulation, and proton-on-target (POT) counting.
These facilities are called \textit{monitored neutrino beams} and can reduce the uncertainty on the flux of $\nu_{e}$ and $\nu_{\mu}$ below 1\%.

ENUBET is the most advanced monitored neutrino beam and all technical challenges have been considered in the design of the ENUBET beamline. It is a conventional beam where protons from the accelerator are sent to a fixed target producing secondary mesons that decay into neutrinos. Two different focusing options are being studied: a purely static focusing system as well as a horn-based transfer line. Several existing accelerator complexes can feed the ENUBET beamline at different proton energies: the 400 GeV (CERN-SPS), the 120 GeV (Fermilab Main Ring), and the 30 GeV (JPARC) protons have all been simulated to estimate the secondary yield with several targets.

\begin{figure}[h]
\centering
\includegraphics[scale=0.45]{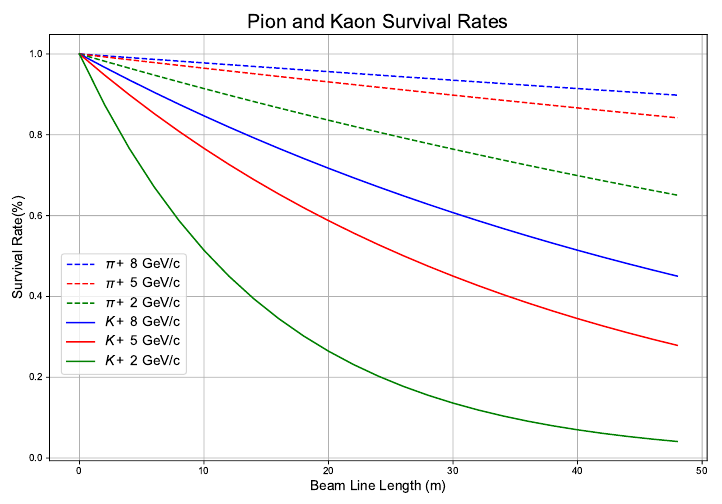}
    \caption{Survival probabilities for low momentum pions and kaons, as a function of the beamline length.}
    \label{fig:survivalprob}
\end{figure}

To cover the region of interest of DUNE, the hadron transfer line design is optimized for the transport of 8.5 GeV/$c$ kaons. The electrons produced in the three-body $K_{e3}$ decays together with the $\nu_{e}$ ($K^{+}\rightarrow e^{+}\pi^{0} \nu_{e}$) are measured in a 40 meters long instrumented decay tunnel. Similarly, ENUBET measures the large-angle muons from the $K^+ \rightarrow \mu^+ \nu_\mu$ decay with a signal to noise ratio larger than 6 and can monitor at single-particle level the muons from pion decays by replacing the hadron dump with a range-meter. A neutrino detector is placed 50 m from the end of the decay tunnel.
The guiding principles for the beamline design are the kaon flux maximization at the tunnel entrance, the reduction of beam size and divergence to contain the hadron beam in the decay volume, and the total length of the transfer line to minimize early kaon decays. Figure~\ref{fig:survivalprob} shows the survival probability of pions and kaons produced in neutrino beams below 10 GeV/$c$ as a function of the length of the beamline.

The static transfer line design makes use of normal-conducting quadrupoles and dipoles: as they run at constant current and do not need to be pulsed, they can be used with second-long proton extractions (slow-extraction). For example, the current slow extraction operation mode at the CERN SPS produces a few second long proton spill for fixed target experiments like ProtoDUNE and it is characterized by low particle rates.  This offers several advantages for monitored beams since pile-up at the instrumented decay volume poses stringent limits on the instantaneous particle rate. Besides, standard focusing magnets are simpler, more robust, and less expensive than horns. This comes at the expense of the flux at the decay tunnel and the lower neutrino yield at the neutrino detector because quadrupole multiplets have smaller acceptance at the ENUBET hadron energies with respect to horns.

\begin{figure}[h]
\centering
\includegraphics[scale=0.8]{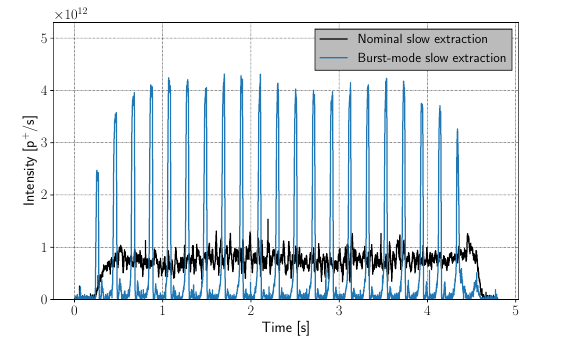}
    \caption{Comparison between a nominal slow extracted spill and a “burst-mode slow extraction” one. The spill profiles have been measured with a secondary emission monitor at the SPS, during dedicated machine tests. The same intensity is extracted in the two cases.}
    \label{fig:burstslowextr}
\end{figure}

For this reason, the ENUBET collaboration is also pursuing a horn-based beamline. To cope with pile-up requirements in the decay region, studies have been carried out at the CERN SPS to combine the need of pulsing the horn with a proton slow extraction scheme. A new pulsed slow extraction scheme (\textit{burst-mode slow extraction}) was successfully developed and tested: 2-to-10 ms proton pulses were repeated at 10 Hz for the full duration of the extraction~\cite{ENUBET_spsc_2020, ENUBET_spsc_2021}.
An example of this burst-mode slow extraction is presented in Fig.~\ref{fig:burstslowextr} and compared with a nominal slowly extracted spill. Note that the same number of protons are extracted in both cases.
The horn design optimization to maximize the flux of focused secondaries in the sought-for momentum range is carried out with a full GEANT4 simulation considering various horn geometries (single and double parabolic, and MiniBooNE-like geometries). It was implemented with a Genetic Algorithm (GA) in a standalone way using a beamline-independent figure-of-merit. Further studies on the dedicated beamline configuration for the horn-focused beam are underway to determine the flux increase with respect to the static option. 

\begin{figure}[h]
\centering
\includegraphics[scale=0.49]{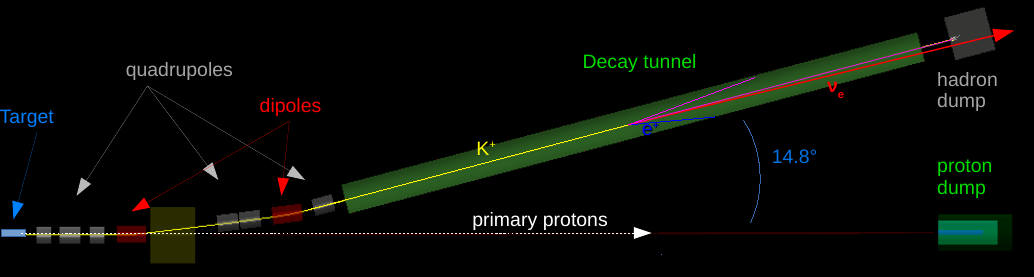}
    \caption{Schematic of the ENUBET static transfer line. Bending elements: dipoles are shown in red while quadrupoles are shown in grey. The hadron dump is placed downstream the tagger exit.}
    \label{fig:tlr6}
\end{figure}

In Figure~\ref{fig:tlr6} we show the most recent static beamline configuration of the ENUBET project: a quadrupole triplet (aperture radius of 15 cm) is placed after the proton target and two identical normal-conducting bending dipoles with a 1.8 T field provide a total bending of the beam with respect to the primary proton line of 14.8°. 
Such a large bending angle suppresses background neutrinos produced by the early decay of neutral kaons in the proximity of the target.
The hadron dump is placed at the tunnel exit. Optics was optimized using the TRANSPORT code while particle transport and interactions are fully simulated with G4Beamline considering 400 GeV/$c$ protons interactions on a graphite target. Proton-graphite interactions were simulated with FLUKA. Doses accumulated in the transfer line are estimated using FLUKA, too. 

\begin{figure}[ht]
\centering
\includegraphics[scale=0.3]{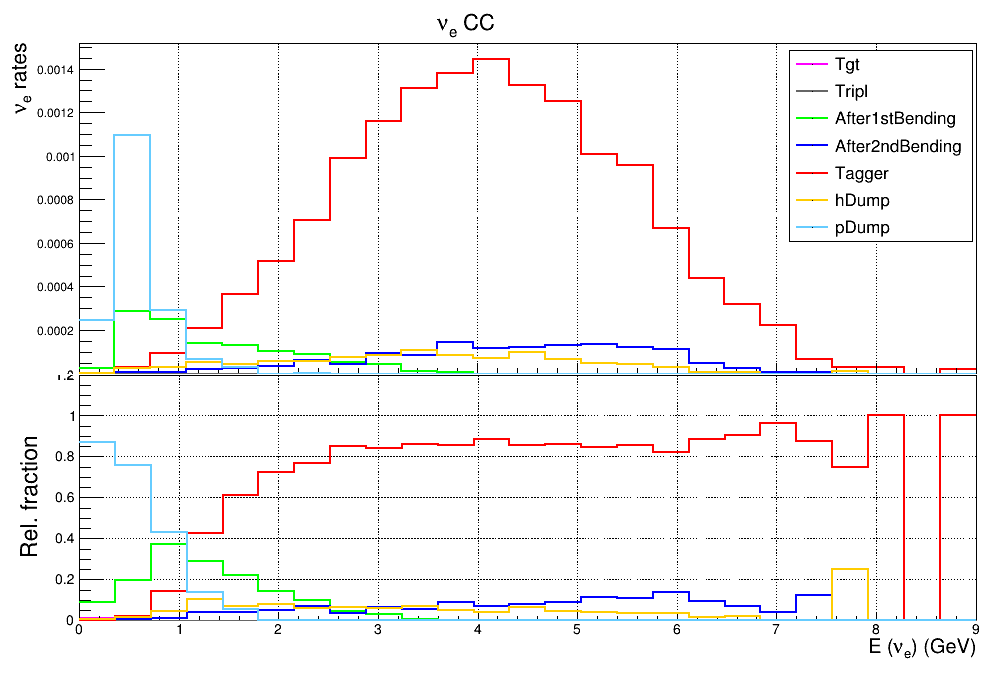}
    \caption{Top: $\nu_{e}$-CC interactions considering a $6\times6$ m$^{2}$front face of a neutrino detector located 50~m downstream the ENUBET decay tunnel exit. The events are divided into categories corresponding to the position along the transfer line where the neutrino was generated. The red spectrum corresponds to neutrinos generated inside the tagger. Bottom: relative fraction of each category to the total $\nu_{e}$-CC rate.}
    \label{fig:nueCCenubet}
\end{figure}

The beamline produces the $\nu_{e}$ CC spectrum shown in Fig.~\ref{fig:nueCCenubet} when considering a   neutrino detector located 50~m downstream the tunnel exit with a 6~$\times$~6~m$^{2}$ size in the plane perpendicular to the beam axis. Classifying the different components of the spectrum based on the neutrino creation point along the transfer line, it is clear that a loose energy cut would already be good enough to separate the monitored component coming from the decay tunnel from other neutrinos. 73.5\% of the total $\nu_{e}$~CC flux is generated inside the tunnel (it exceeds 80\% considering only energies above 1 GeV). By assuming $4.5 \times 10^{19}$ POT/year at the SPS, 10$^{4}$ $\nu_{e}$~CC are collected in a 500-ton neutrino detector (e.g. ProtoDUNE~\cite{DUNE:2020cqd} or ICARUS~\cite{MicroBooNE:2015bmn,ICARUS:2004wqc}) in about 2 years.
Below 1 GeV, the main component is produced in the proton-dump region, which can be further suppressed by optimizing the proton dump design and position. The remaining 12\% of the flux is given by the straight section in front of the tagger plus the decays or interactions occurring after the tunnel and in the hadron dump. Its contribution to the flux is corrected relying on simulation and particle monitoring along the beamline.
A static line design for a monitored neutrino beam would also pave the way to the so-called \textit{tagged neutrino beams}~\cite{Hand1969,bernstein,Pontevcorvo1979} that will be briefly discussed later.

\begin{figure}[h]
\centering
\includegraphics[scale=0.5]{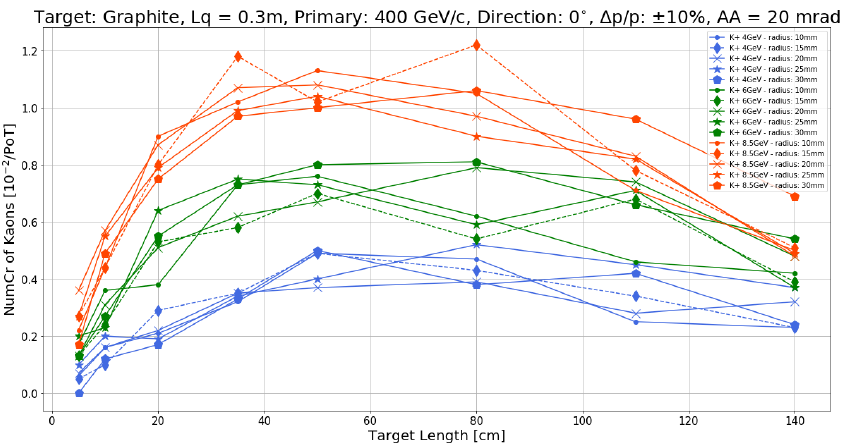}
    \caption{Kaon yields as a function of the graphite target length for a 400 GeV/c proton beam. Colors refer to different kaon momenta while the marker style identifies the target radius.}
    \label{fig:targets}
\end{figure}

Detailed studies were performed by ENUBET also on the target, hadron, and proton dumps. The most common materials chosen for targets at proton accelerators are low-Z, radiation hard materials. ENUBET performed optimization studies on graphite ($\rho$=2.2 g/cm$^{3}$), beryllium ($\rho$=1.81 g/cm$^{3}$) and Inconel ($\rho$=8.2 g/cm$^{3}$). Inconel -- an austenitic nickel-chromium-based superalloy --  is a novel target choice currently considered also by the nuSTORM collaboration (see below). Each target was modeled as a cylinder with a variable radius between 10 to 30~mm and with a different length from 5 to 140~cm. Figure~\ref{fig:targets} shows the kaon yields obtained for graphite targets. The figure-of-merit employed for the optimization is the number of kaons with a $\Delta p/p=\pm 10\%$ that enter an angular acceptance of $\pm$20~mrad. The most promising materials turned out to be graphite and Inconel-718. 400 GeV/$c$ protons impinging on a graphite target 70 cm long and with a radius of 3 cm were thus used for the static beamline simulation, while the Inconel option is considered quite promising for the horn beamline: the small loss of kaon flux is compensated by the larger acceptance of the horn. Moreover, Inconel produces fewer positrons reducing the expected background in the decay tunnel.

The positron flux coming directly from the target region that hits the instrumented decay tunnel walls represents the most critical background since it cannot be rejected by particle identification. A tungsten thin foil placed directly downstream the target reduces this background by orders of magnitude and was optimized comparing the amount of background expected with respect to the kaon flux for different thicknesses of the foil.
The proton dump is made by a 3 m long graphite core, surrounded by aluminum and covered by iron.
This design is inherited by several dumps of high-intensity proton beams operated at CERN. Its final design and position will be further studied to reduce the proton dump contribution to the neutrinos recorded in the neutrino detector. The structure of the hadron dump is similar. Its optimization, however, is aimed at reducing the back-scattering. Back-scattering results in particles hitting the tunnel walls increasing the background and in the growth of neutron fluence that contributes to the total dose received by the instrumentation. The hadron dump cannot be placed too far away from the tunnel exit to minimize kaon decays in a region where the leptons cannot be monitored.
A design that mitigates back-scattering events consists of a hadron dump placed two meters after the decay tunnel with a graphite core (50 cm diameter), inside a layer of iron (1 m diameter), covered by borated concrete (4 m diameter). In addition, 1 m of borated concrete is placed in front of the hadron dump leaving an opening for the beam. This has provided a significant reduction of the flux coming from the dump all along the tagger, in particular in the last few meters where the neutron fluence is particularly relevant.

\begin{figure}[h]
\centering
\includegraphics[scale=0.3]{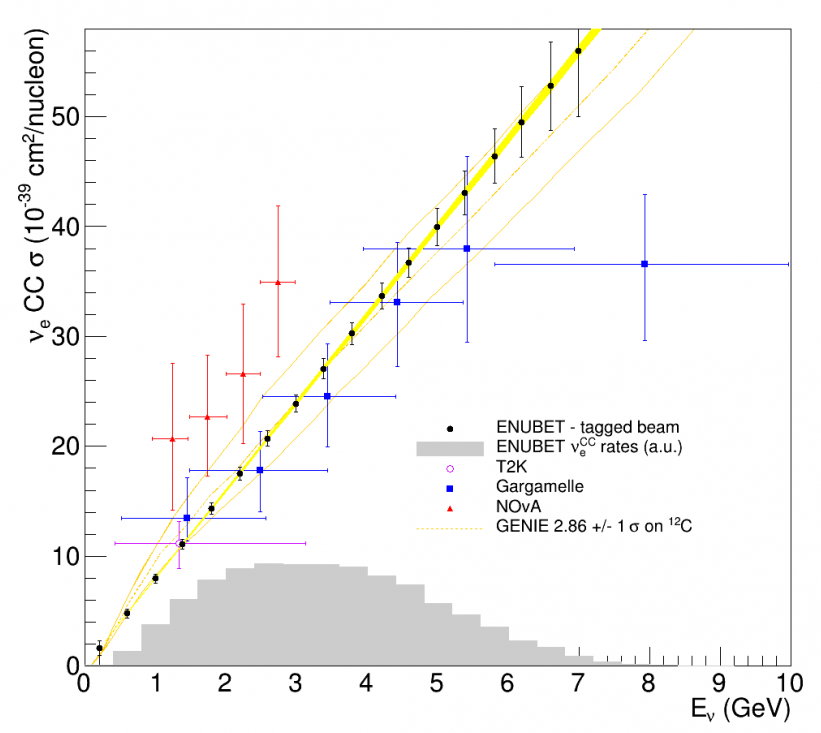}
    \caption{Impact of the ENUBET data on existing $\nu_{e}$ cross-section measurements assuming 1\% flux precision.}
    \label{fig:xsecenubet}
\end{figure}

The impact of the ENUBET data on the $\nu_{e}$ cross-section measurement assuming 1\% flux precision is shown in Figure~\ref{fig:xsecenubet} and compared with current data.

\begin{figure}[h]
\centering
\includegraphics[scale=0.47]{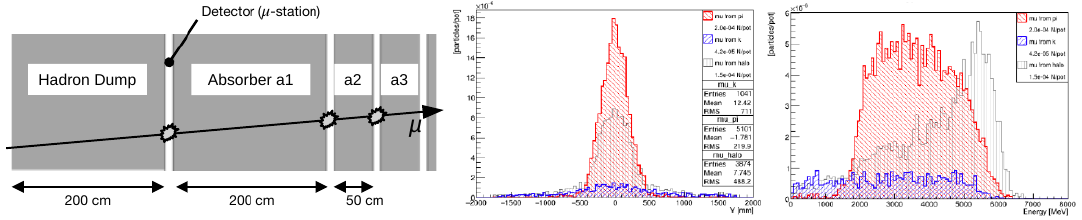}
    \caption{Left: Schematic of the muon stations and absorbers configuration to be installed at the end of the ENUBET tagger calorimeter. The grey slabs represent the absorbers (made out of iron or rock) while the white slices, 8 in total, are the muon detector planes. Center: spatial distribution along one of the two dimensions of a detector layer. Right: energy distribution. Red and blue are the signal contribution, from pion and kaon decays. The gray distribution represents halo-muons.}
    \label{fig:mustations}
\end{figure}

The results obtained by ENUBET fostered the possibility to constrain the low-energy $\nu_{\mu}$ flux from pion decays by monitoring associated muons emitted at low angles by instrumenting the hadron dump. Muons that exit the decay tunnel and go through the hadron dump can be measured by detectors placed between absorber layers. With a static focusing beamline and fast muon detectors, it is possible to reconstruct muons on an event-by-event basis and measure their momentum from their range in the detector stations. The instrumented hadron dump allows to separate the three different components of the muon spectrum: as shown in Figure~\ref{fig:mustations}, the measured momentum can be exploited to separate muons from pions, and beam-halo muons from muons of $K_{\mu 2}$ ($K^+ \rightarrow \mu^+ \nu_\mu$) and $K_{\mu 3}$ decays that, in turn, can be used to determine the $\nu_{\mu}$ flux. The energy of $\nu_{\mu}$ can be studied in bins corresponding to muon momenta reconstructed by range as they show a clear anti-correlation due to the 2-body decay kinematics, allowing to constrain the $\nu_{\mu}$ shape and normalization.
The ENUBET collaboration is studying a  system of 8 muon stations to be placed right after the hadron dump with iron absorbers from 2 m (upstream) to 0.5 m (downstream) depth (see Fig.~\ref{fig:mustations}). Muon and neutron fluences have already  been estimated: the most upstream detector needs to cope with a muon rate of 2 MHz/cm$^{2}$ and the total neutron fluence integrated over the experiment lifetime is 10$^{12}$ 1-MeV-eq/cm$^{2}$. The radiation damage expected in a monitored beam is much smaller compared to beams where the dump is used to stop non-interacting primary protons. This is the reason while instrumented dumps are not an option in conventional beams like the ones used for long-baseline experiments.

It is worth mentioning that ENUBET is working also on a site-dependent multi-momentum beamline to cover the lower energy region, which is of interest for HK: the current design of the multi-momentum beamline uses existing CERN magnets. The beamline optics is flexible enough to select hadron momenta down to 4 GeV or less. The design consists of quadrupoles for focusing and two dipoles for momentum selection and a large total bending to ensure the separation of the different $\nu_{e}$ components at the neutrino detector. 

The possibility of measuring the flux of $\nu_\mu$ with a precision comparable with $\nu_e$ is a significant asset for ENUBET. Since ENUBET is a narrow band beam and the pions produce muon-neutrinos in a two-body decay, there is a strong correlation between the neutrino angle and its energy. As a consequence, the measurement of the interaction vertex of the neutrino in the detector (and, hence, its angle) provides a measurement of the energy with an average precision of 10\%. It is worth stressing that this indirect energy measurement (\textit{narrow-band off-axis technique}) can be performed at the level of a single neutrino without resorting to the energy reconstruction at the detector. This is the main tool to remove all reconstruction biases mentioned in Sec. \ref{sec:future_ND} and suppress the systematics on neutrino interaction modeling that plague the ND measurements without any tuning with data.


\subsection{Tagged neutrino beams}
As anticipated, a monitored neutrino beam with a slow proton extraction scheme would open the possibility for a time-tagged neutrino beam where the neutrino recorded in the detector is associated in time with the observation of the lepton from the parent hadron in the decay tunnel. Tagged beams require a detector system with a time resolution of $\mathcal{O}$(~100 ps). Reaching this level of accuracy in the decay volume would represent another prominent asset. A precise time coincidence between the reconstructed lepton and the neutrino interaction provides the neutrino flavor without looking at the lepton in the neutrino detector produced by CC interactions. Besides, if the other products of the decays (of $\pi$ or K) are measured, the energy of the neutrino can be constrained with even higher precision than a monitored beam.
The neutrino sample selected with a time-tagged beam would have an unprecedented purity. 
For every neutrino interaction in the detector, one has to select the lepton candidates in the decay volume compatible in time with the neutrino within the detector resolution. If the neutrino interaction time is $t_{\nu}$ and the lepton is time-tagged at $t_{l}$, considering the distances from the decay point  of the two interactions ($d_{\nu}$ is the decay-neutrino interaction distance and $d_{l}$ is the decay-lepton interaction distance), we can write:
\begin{equation}
t_{\nu}-t_{l}=d_{\nu}-d_{l}\frac{E_{l}}{p_{l}} 
\end{equation}
The neutrino production point is unknown so $d_{\nu}-d_{l} \frac{E_{l}}{p_{l}}$ has to be approximated by $z_{\nu}-z_{l}$, the distance of the neutrino interaction vertex from the lepton impact point in the decay volume projected along the axis of the decay volume. This approximation introduces a time spread due to the fact that the neutrino and the lepton are not exactly collinear. This irreducible spread ($\delta_{irr}$) is about 100 ps and sets the bar of the timing precision needed in a tagged beam. The time match requires:
\begin{equation}
\Delta t = \left| t_{\nu}-t_{l}-(z_{\nu}-z_{l}/c)\right| < 
 \sqrt{ \left(\delta_{\nu  tagger} \right)^2 +  \left(\delta_{\nu  Det} \right)^2 + \left(\delta_{irr} \right)^2}
\end{equation}

where the time resolution of the lepton tagger in the decay tunnel and the neutrino detector are $\delta_{tagger}$ and $\delta_{\nu  Det}$, respectively.
The main background for a tagged beam is given by accidental coincidences. They are proportional to the combined time resolution $\delta = \sqrt{ (\delta_{tagger})^2 + (\delta_{\nu Det})^2}$, and the time match $\delta$ must be smaller than $\delta_{irr}$ to achieve a high purity sample.
The number of true time-coincidences can also be improved by increasing the geometrical acceptance of the detectors in the decay volume, which implies that the technology used for these detectors must provide excellent timing over large areas.
A dedicated project, NUTECH~\cite{NUTECH}, is exploring different technologies for fast timing in large areas like Cherenkov Micromegas (PICOSEC~\cite{PICOSEC}), Large Areas Picosecond PhotoDetectors (LAPPD~\cite{LAPPD}) or LYSO(Ce) crystals. In fact, even the current photon veto of ENUBET, which is based on standard plastic scintillators, has achieved a 200 ps resolution and further improvements are envisaged to reach the 100 ps level. The same precision can be achieved by the neutrino detectors thanks to the Cherenkov signal of water-based detectors and the 128~nm scintillation light of argon.

\subsection{Muon beams}
Moving away from conventional neutrino beams, a precise flux measurement can be performed through the muon decays $\mu^{\pm} \rightarrow e^{\pm} + \nue (\nubare) + \nubarmu (\numu)$ by storing and accelerating muons in a ring. This is the idea behind the Neutrino Factories~\cite{Geer:1997iz,DeRujula:1998umv,Choubey:2011zzq} where the beam composition is precisely known: 50\% $\nue (\nubare)$ and 50\% $\nubarmu (\numu)$. There are three main advantages in this approach: first, the number of muons in the storage ring can be easily measured providing an excellent estimate of the flux. Then, as neutrino flavors are determined purely by the muon decay, the beam is not contaminated by neutrinos originating from other particles. Finally, a Neutrino Factory offers a powerful source of $\nue(\nubare)$ unlike conventional beam (including ENUBET) where they never exceed 10\% of the total flux. Building this kind of facility presents, however, major technical challenges that have not been superseded despite decades of R\&D. Muons must be accelerated fast with a reduced transverse momentum $p_{T}$  before injecting them in the storage ring. A muon-based neutrino beam (and, even more, a muon collider) thus requires a strong muon cooling. Some encouraging results were recently achieved by  the analysis of the MICE data~\cite{MICE:2019jkl}. Still, the first full-fledged muon storage ring has not been built, yet. It has been designed by the nuSTORM collaboration~\cite{Adey:2013pio} as a powerful facility for neutrino cross sections.

\begin{figure}[h]
\centering
\includegraphics[scale=0.59]{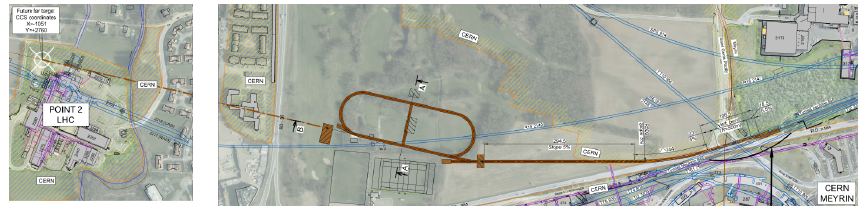}
    \caption{Overview of the implementation of nuSTORM at CERN. The nuSTORM ring and associated infrastructure is shown in the right panel. The possible location for a far detector is shown in the left panel.}
    \label{fig:nustorm}
\end{figure}

Even if nuSTORM is significantly more complex than a monitored neutrino beam, it offers much larger statistics of $\nu_e$ and can feed an ambitious cross section measurement program.
Figure~\ref{fig:nustorm} shows a schematics of the nuSTORM facility. nuSTORM makes use of a horn to focus pions that are then transported by a series of quadrupoles and dipoles to the injection point of the storage ring. The ring is composed of several FODO cells: a large acceptance focusing and defocusing quadrupole lattice. The circular ring has two long straight sections where muons of $E_{\mu} \sim 5$ GeV from pion decays produce neutrinos of both electron and muon flavors. 

The three-body kinematics of muon decay does not allow for the measurement of the neutrino energy at the single-particle level that can be employed by ENUBET through the narrow-band off-axis technique. Still, nuSTORM can achieve excellent precision on the fluxes operating an FBCT (Fast Beam Current Transformers) in the decay ring to measure the muon intensity. These devices are based on toroids and developed at CERN, and are needed for fast extractions where muons are bunched in tens of $\mu$s up to a few ms. 

A feasibility study for nuSTORM at CERN~\cite{nuSTORM_feas} reports that the experiment would use protons extracted from the SPS at 100 GeV and direct them on a low-Z target placed inside a focusing horn. After the horn, particles are collected by a pair of quadrupoles. Then the beam goes through a short transfer line composed of dipoles, collimators, and quadrupoles. The detailed design of the target and capture system has to cope with radiation safety and full containment in the transport of a beam with a momentum spread of $\pm$10\%. To fulfill these requirements the experiment is considering the scheme that has been successfully used in the CERN PS complex at the Antiproton Decelerator as well as the work done for the CENF and LAGUNA-LBNO projects for the target-horn system.

For the pion transfer, modular construction with simple quadrupole FODO cells and achromatic dipoles was chosen to ensure more flexibility in the current design phase. The design of the proton absorber will be based on already existing SPS internal dumps. The first achromatic bending section bends particles in the momentum bite of the beamline towards the ring and far from the proton absorber. A quadrupole FODO lattice is used to transport the beam to the second achromatic bending section which is followed by a beta-function matching before injecting the beam into the ring. The length of this section is short enough to transport also low momentum pions.

\begin{figure}[h]
\includegraphics[scale=0.58]{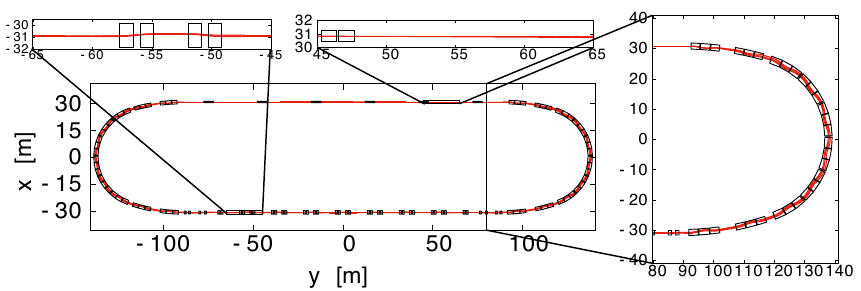}
    \caption{Schematic drawing of the hybrid concept adopted for the nuSTORM muon storage ring. The beam circulates in an anticlockwise direction. The production straight (at z$\sim$30 m) is composed of large aperture quadrupoles that produce the large values of the betatron function required to minimize the divergence of the neutrino beam produced in muon decay. The lattices of the arcs and return straight are based on the FFA concept  and allow a large dynamic aperture to be maintained.}
    \label{fig:nustormring}
\end{figure}

The nuSTORM decay ring is a storage ring with a circumference of $\sim$616 m made by large apertures magnets (see Fig.~\ref{fig:nustormring}). In addition, strong bending magnets are used in the arcs to minimize the arc length and maximize the number of useful muon decays. The nuSTORM beamline was recently redesigned to serve the neutrino-scattering program and store muons in the 1-6 GeV momentum range with an acceptance of $\pm$16\% to increase the neutrino flux. A hybrid concept was then adopted in order to simultaneously have large momentum acceptance and high muon accumulation efficiency: the conventional FODO optics of the production straight section are combined with Fixed-Field Alternating gradient cells (FFA) with zero chromaticity in the arcs and the return straight.

The magnets employed in different sections are: superconducting combined-function magnets with fields up to $\sim$2.6 T in the arcs, combined-function room temperature magnets in the return straight and large aperture room temperature quadrupoles in the production straight. In both the production and return straight the mean betatron function contribution to the angular spread of the neutrino beam is minimized so that both can be used to serve a neutrino physics program. Finally, the arcs are connected with the injection and return straights using specific matching sections. At the end of the production straight undecayed pions and muons outside the selected momentum will be directed to a dump. The Near Detector is located 50~m downstream of the dump in a shallow experimental hall. In the CERN implementation of nuSTORM, the neutrino beam is directed to LHC Point 2 where it is possible to install an additional neutrino detector at a larger distance (see Fig.~\ref{fig:nustorm}). 

To evaluate the performance of the hybrid design for the storage ring the nuSTORM collaboration performed tracking studies that showed that the dynamical acceptance of the machine is about 1 $\pi$~mm~rad in both transverse planes matching the requirements of the experimental program.
If implemented at CERN, an upgrade of the present extraction kickers is needed to work with two 10.5 $\mu$s proton pulses, and approximately 165~kW of beam power will be delivered in a cycle of 3.6 s corresponding to a total intensity of $4\times10^{13}$ per cycle.

The feasibility of executing nuSTORM at CERN is carried out in the framework of Physics Beyond Colliders, which has highlighted interesting synergies with the ENUBET project. In particular, three main common areas have been identified: the target facility, the first stage of the meson focusing, and the proton dump. The two collaborations are currently evaluating this possible synergy at the facility level.

\section{A new generation of neutrino-beam detectors}
\label{sec:precision_detectors}
High-precision neutrino beams, as the NP06/ENUBET and nuSTORM facilities described in the previous section, allow to disentangle the flux term and to reduce the corresponding systematics in Eq.~\ref{eq:near}. 
In order to retrieve precise information on the other two terms, the detector efficiency and the neutrino cross section, a facility based on different neutrino detection techniques is highly desirable.
In particular a program able to boost the physics reach of DUNE and HK by studying neutrino interactions in Argon and water, but at the same time capable of providing high precision data on neutrino interactions with low-Z targets should be envisaged.

Liquid argon TPC prototypes of moderate mass (protoDUNE-SP and pro\-to\-DUNE-DP~\cite{DeBonis:2014jlo}) have been installed and operated with charged particle beams in EHN1 at CERN, with the main goal of demonstrating the detector technology of the DUNE far detector. If protoDUNE-DP is based on a more innovative concept, with a gaseous argon layer above the liquid phase used to amplify drifting charges for a better signal-to-noise ratio, protoDUNE-SP stands on the single-phase liquid argon technology, successfully operated by ICARUS~\cite{ICARUS:2004wqc} and MicroBooNE~\cite{MicroBooNE:2016pwy} and used by the upcoming SBND detector~\cite{McConkey:2017dsv}, in the Fermilab Short-Baseline Neutrino program, and it has already achieved excellent performance~\cite{DUNE:2020cqd}. 

\begin{figure}[ht]
\begin{center}
\includegraphics[scale=0.85]{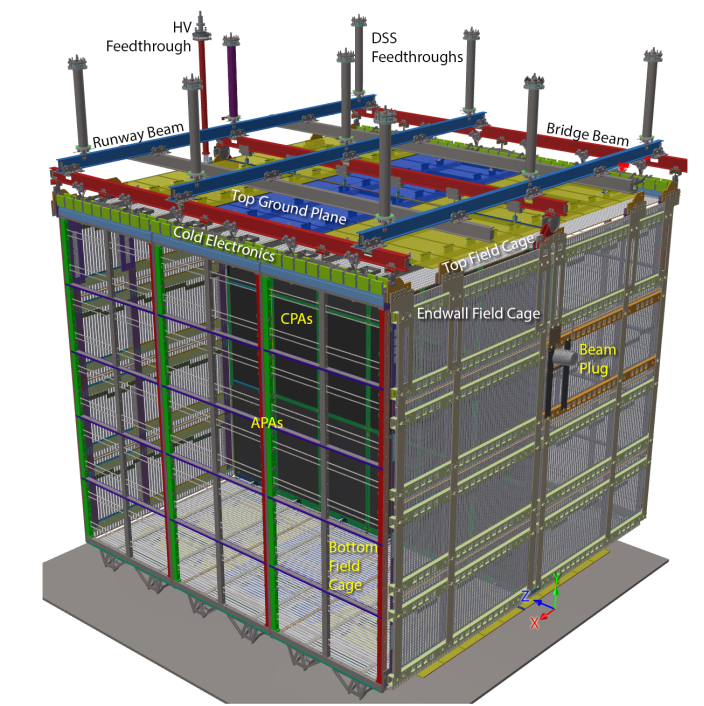}%
\includegraphics[scale=0.85]{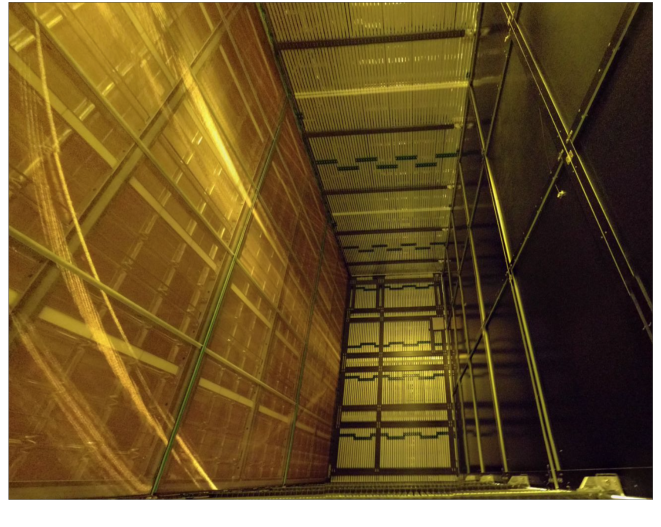}
\caption{Left: Overview of the protoDUNE-SP detector. Right: a photo of one of the two drift volumes with the anode plane embedding sense wires and photodetection systems on the left and the cathode plane on the right.}
\label{fig:protoDUNE}
\end{center}
\end{figure}

The active volume of protoDUNE-SP extends for 7~m along the beam direction, 7.2~m in the drift direction and 6.1~m in height, for a total mass of about 400~tons (Fig.~\ref{fig:protoDUNE}). The detector is divided into two drift regions by a cathode plane in the middle and the readout is performed on each side by sense wires placed on anode planes. The anode planes embed photon detectors to collect scintillation light from ionized LAr used to provide the absolute timing of the neutrino interaction and an independent measurement of the deposited energy.

Unlike the liquid argon detector of the DUNE ND complex (ND-LAr~\cite{AbedAbud:2021hpb} - see Sec.~\ref{sec:future_ND}), that is composed by 35 small optically separated TPC modules to withstand the large neutrino interaction rate at the near location, the large size of the protoDUNE TPC allows for an almost full containment of the neutrino interaction. Moreover, it is based on the same components and has the same maximum drift length as the first DUNE far detector module. These features make protoDUNE-SP an ideal detector for the precise determination of the $\sigma(E)\times\epsilon(E)$ product for DUNE, provided that the neutrino flux is known at the percent level, like in ENUBET and nuSTORM.

To decouple the neutrino cross section in argon from the detector efficiency, the simultaneous use of a liquid and gas phase TPC would be an ideal solution.
Atmospheric pressure argon TPCs have been used in neutrino experiments, like the ND280 near detector of T2K and its upcoming upgrade~\cite{Abe:2019whr}. A large volume TPC~\cite{ALICE:2000jwd} is used in the ALICE experiment at LHC. The TPC employed in the ND-GAr detector~\cite{AbedAbud:2021hpb} of the ND complex of DUNE will be based on the same design of ALICE, but will be operated at a ten times bigger pressure to enhance the amount of target nuclei in the volume and, hence, the neutrino event rate. The active volume is divided into two parts by a central cathode and has a radius of 2.6~m and a length of 5~m, for a total mass of 1~ton. The endplates are instrumented with chambers equipped with MWPCs and pad planes.

The main advantages of a high pressure gaseous TPC with respect to a liquid argon TPC are the superior momentum resolution given by track bending in the TPC magnetic field, a better particle ID, particularly for proton-pion separation, and a significant lower threshold that allows the reconstruction of low-kinetic-energy protons and pions. The full reconstruction of the hadronic system is of paramount importance for the evaluation of final state interactions and for neutrino-nucleus interaction modeling. As an example, ND-GAr will be able to reconstruct protons with a kinetic energy above 20~MeV with an efficiency exceeding 90\% and preliminary studies of a machine learning based reconstruction show that the threshold could be pushed further down to 5~MeV~\cite{AbedAbud:2021hpb}.

\begin{figure}[ht]
\begin{center}
\includegraphics[scale=0.92]{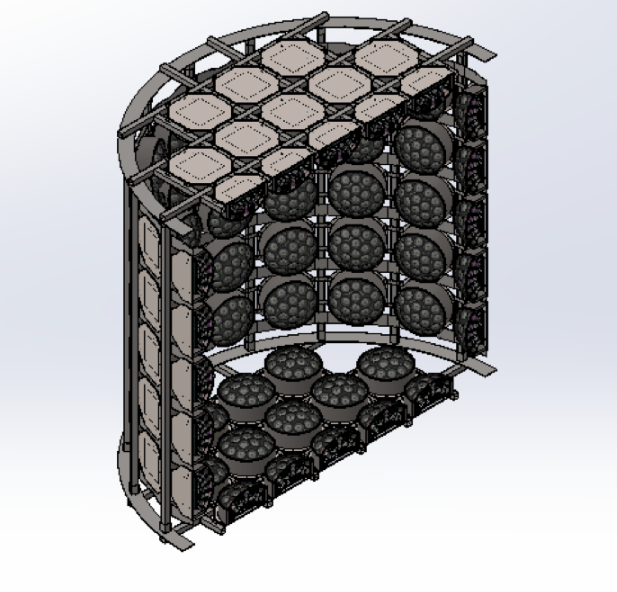}%
\includegraphics[scale=0.75]{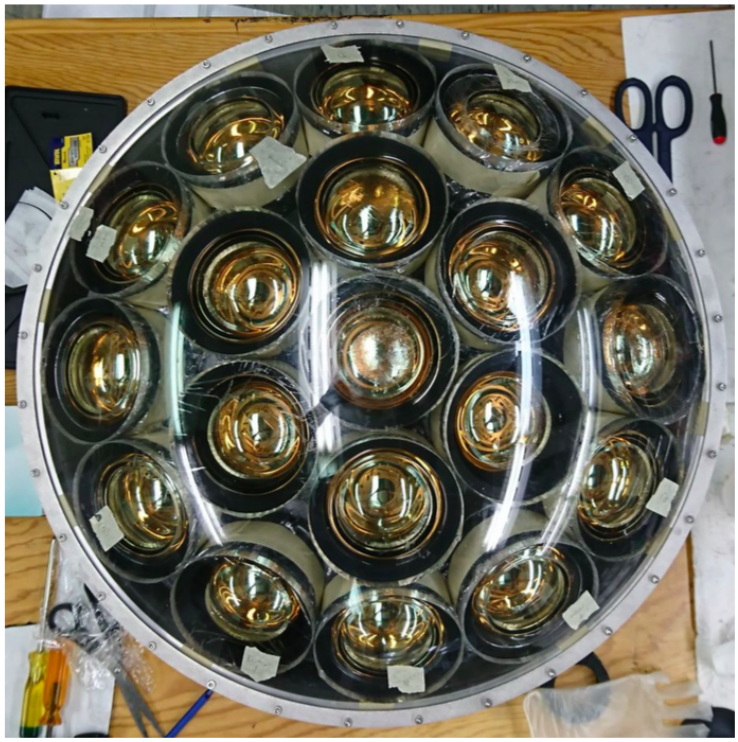}
\caption{Left: Cross section drawing of WCTE water tank with the multi-PMT layout inside the detector. Right: Top view of a multi-PMT prototype.}
\label{fig:WCTE}
\end{center}
\end{figure}

Concerning the cross section measurement in water, an interesting opportunity is represented by the proposed Water Cherenkov Test Beam Experiment (WCTE)~\cite{WCTE:2021prop} that will be exposed to charged particle beams at CERN in 2022 or 2023. The setup is based on a Water Cherenkov tank of 4~m height and 4.1~m radius, for a total mass of $\sim$50~tons and the light will be collected by 19 small (8 cm diamter) photomultipliers arranged in 128 multi-PMT optical modules (Fig.~\ref{fig:WCTE}). The photosensor size is driven by the much smaller dimensions of the detector: Cherenkov rings will cover a reduced area and the event reconstruction would be deteriorated by using the same 50 cm diameter PMTs as SK and HK. The same arrangement will be used in the Intermediate Water Cherenkov Detector (IWCD) of HK.

Like protoDUNE, a neutrino exposure of the WCTE tank would profit of the characterization of the detector response by charged particles to be performed at the CERN East Area in 2022-23. Moreover, the detector technology and the event reconstruction are very similar to those of the HK far detector and allow for a reduction in the systematic uncertainties budget in cross section and oscillation analyses.
Given the small fiducial mass of the detector we do not expect to perform a statistically significant measurement with $\nu_e$ in a moderate intensity beam like ENUBET, but $\nu_{\mu}^{CC}$ interactions will provide a large sample for the disentanglement of detector effects and double-differential cross section measurements.
The detector size also limits the muon containment and a downstream spectrometer must be envisaged for momentum measurements.

Disentangling the neutrino cross section in water from the detector efficiency can be accomplished by using fine-grained detectors with water targets.
The WAGASCI experiment~\cite{Giorgio:2019iuq} exploits a 0.6 ton water target encompassed by a 3D grid-like structure of plastic scintillator strips enclosing cells of $\mathcal{O}(cm)$ linear size. The setup is complemented by two side modules with steel plates interleaved with scintillator slabs and a downstream magnetised detector equipped with plastic scinitillators (BabyMIND) for muon momentum measurement.
An alternative option has been developed by the NINJA collaboration~\cite{NINJA:2020gbg}: nuclear emulsion films and iron plates (500~$\mu$m thick) are interleaved in a sandwich like structure with 2~mm thick water layers and a downstream scintillating fiber tracker is used to timestamp and match tracks in the emulsions. With this setup the momentum threshold of hadrons generated in the neutrino interaction can be pushed down to 200~MeV/c and 50~MeV/c for protons and pions respectively.

The precision measurement of neutrino scattering off hydrogen and deuterium (the lowest Z isoscalar nucleus) is pivotal to build reliable cross section models not affected by nuclear effects, that can be extrapolated to higher Z materials, and to perform detailed studies on the nucleon structure exploiting a bare weak probe.
Indirect approaches to measure the $\nu$-H cross sections in hydrocarbon targets exploiting the transverse momentum imbalance or subtracting measurements on thin dedicated graphite targets have been already discussed in Sec.~\ref{sec:future_ND}.
A straightforward way to perform the measurement is the direct use of a liquid hydrogen/deuterium target. In the past, bubble chambers were extensively used in this way: the Argonne 12' bubble chamber with few GeV neutrinos on H and D~\cite{Barish:1978pj}, the series of experiments performed at higher energies at the Big European Bubble Chamber (e.g.~\cite{Allasia:1990uy}), and the Fermilab 15' bubble chamber~\cite{Kitagaki:1983px}.
The proponents of~\cite{snowmass-H} suggest to re-use the Fermilab chamber for an upgraded experiment with liquid hydrogen that could profit of modern digital camera technology and machine-assisted reconstruction techniques to improve precision and data analysis speed. On the other hand, the statistics is limited by the slow operation cycle of the chamber, and auxiliary detectors are needed for event containment due to the long radiation and interaction length of hydrogen.


\section{A graded strategy toward high-precision neutrino physics}
\label{sec:conclusions}

At the conclusion of this paper, we recap the main motivations and advances toward a new generation of cross section experiments. The results shown in Sec.~\ref{sec:current_generation} testify for the need of high-precision experiments:

\begin{itemize}
    \item even in the most precise measurements (see for instance Fig.~\ref{fig:double_differential}) the flux uncertainty contribute to a large fraction of the systematic uncertainty. Double-differential cross section measurements are not yet systematically limited but the DUNE and HK near detectors of Sec.~\ref{sec:future_ND} will soon reach this limit. A per-cent knowledge of the flux is, then, mandatory to reduce the systematic uncertainty once statistical errors will be lower than 10\%. The  use of standard candles mitigates this issue but the most desirable solution is a flux measurement independent of the neutrino detector.
    \item the use of non-monochromatic beams is the root of the model dependence of cross section results. Major advances have been achieved once experimenters became aware of the theory priors implicit in the definition of QE, RES, and DIS events (see for instance Sec.~\ref{sec:qe_axial_mass}). Still, the use of a narrow band beam is mandatory to decouple interaction modeling from experimental data and reduce the systematic budget due to the energy-integrated flux and the bias coming from energy reconstruction. The PRISM technique is an important mitigation tool. Once more, the ideal solution is a narrow band beam where the neutrino energy is measured with $\simeq  10$\% precision independently from the detector reconstruction.     
    \item the decoupling of cross-section and detector efficiency is a major experimental challenge. It is not strictly necessary for long-baseline experiments but it plagues the interpretation of experimental data for model building. Decoupling can be achieved by combining high- and low-density detectors using the same nuclei. On the other hand, the construction of realistic models of neutrino interactions requires a deeper understanding of low-Z nuclear effects. The use of hydrogen or deuterated materials would be a major asset for electroweak nuclear physics and the study of nuclear media and, in turn, would impact the precision of oscillation measurements.
    \item the construction of an intense \nue beam is extremely valuable to test lepton universality and ground the $\numunue$ and \nubarmunubare oscillation measurements. Beams like ENUBET can provide \nue measurements at the per-cent level, which fulfill the needs of DUNE and HK, but double-differential \nue CC measurements require novel muon beams like nuSTORM.
\end{itemize}

The technology for high-precision facilities is well-grounded for conventional beams like ENUBET, while requires a significant R\& D for muon beams. Similarly, neutrino detectors in liquid and gas phase are available for argon and can be envisaged for water. On the other hand, low-Z measurements are the most direct path toward the understanding of nuclear effects, and low-Z detectors should be reconsidered in a modern perspective. 

While the R\&D for high-precision facilities has achieved remarkable results, a complete site-dependent study is still lacking. It is fortunate that the beam power needed for these facilities is well within the reach of existing proton synchrotrons at CERN, FNAL, and JPARC. A site-dependent study should include a careful evaluation of the impact of a high-precision neutrino beam in the laboratories hosting the proton driver, the corresponding infrastructures, and the engineering assessment, which are mandatory for a reliable estimate of costs. At the time of writing, the most advanced site-dependent study is the one being pursued at CERN for ENUBET and nuSTORM. The CERN infrastructure, accelerators, and experimental halls, like EHN1 serving ProtoDUNE-SP and ProtoDUNE-DP, are ideal to host such a facility and, in a timescale comparable with DUNE and HK, CERN has the opportunity to be the main hub for precision neutrino physics. The laboratories hosting LBNF for DUNE (FNAL) and the HK neutrino beam (JPARC) are natural candidates too. A site-dependent study for FNAL and JPARC is highly desirable since precision experiments do not interfere with the running of long baseline experiments given the limited power needed for short baseline neutrino detectors.
The ideas and proposals discussed in this paper are currently under investigation at CERN in the framework of Physics Beyond collider and in the US within the Snowmass community planning exercise~\cite{snowmass_2021}




\vspace{6pt}

\authorcontributions{Secs.~\ref{sec:introduction} and \ref{sec:conclusions} were drafted by A.B., A.L., and F.T. and reviewed by all authors. G.B. and F.P. contributed to Secs.~\ref{sec:precision_beams} and \ref{sec:precision_detectors}, respectively. Secs.~\ref{sec:current_generation} and \ref{sec:future_ND} were drafted by A.B., M.M., and F.T. Furthermore, M.M. reviewed the status of model building and theory challenges for the whole paper. All authors contributed to the final revision of the manuscript.}

\funding{This work has been supported by the European Union’s Horizon 2020 Research and Innovation
Programme under Grant Agreement no. 681647 and no. 777419, and by the Italian Ministry for Education and Research through NUTECH
(MIUR, bando FARE, progetto NUTECH) and 2017-NAZ-0444 (MIUR bando PRIN2017, CUP H45J17000460006). Tagged neutrino beam studies are also supported by the Dep. of Physics of the Univ. of Milano-Bicocca under the 2018-CONT-0128 project. }


\acknowledgments{We are grateful to our colleagues of the DUNE, ENUBET, HK, and T2K collaborations. Special thanks go to A. Bross, N. Charitonidis, L. Ludovici, and M. Mezzetto for many enlightening discussions on the challenges of neutrino cross section measurements.}

\conflictsofinterest{The authors declare no conflict of interest.} 


\appendixtitles{no} 

\end{paracol}  
\reftitle{References}


\externalbibliography{yes}
\bibliography{bibliography.bib,bib_marco.bib}

\end{document}